\documentclass[lettersize,journal]{IEEEtran}
\usepackage{amsmath,amssymb,amsfonts}
\usepackage{xcolor}
\usepackage{algorithmic}
\usepackage{graphicx}
\usepackage{array} 
\usepackage{booktabs}
\usepackage{multirow}
\usepackage{subcaption}
\usepackage{algorithm,algorithmic}
\usepackage{hyperref}
\hypersetup{hidelinks=true}
\usepackage{textcomp}
\usepackage{cite}
\usepackage[switch,pagewise]{lineno}  
\usepackage[utf8]{inputenc} 

\newcommand{\revised}[1]{\textcolor{black}{#1}}

\begin{document}
\bstctlcite{IEEEexample:BSTcontrol}
\title{Noncontact Multi-Point Vital Sign Monitoring \\with mmWave MIMO Radar}
\author{Wei Ren, \IEEEmembership{Member, IEEE}, Jiannong Cao, \IEEEmembership{Fellow, IEEE}, Huansheng Yi, \\
Kaiyue Hou, Miaoyang Hu, Jianqi Wang, and Fugui Qi, \IEEEmembership{Member, IEEE}
\thanks{This manuscript was submitted on September 13, 2024. This work was supported by 
National Natural Science Foundation of China (62201578),
TRS-RGC Theme-based Research
Scheme (T41-603/20-R), and Joint Founding Projectof Innovation Research Institute, Xijing Hospital. 
\textit{(Corresponding author: Fugui Qi.)}
}
\thanks{Wei Ren and Jiannong Cao are with the Department of Computing, The Hong Kong Polytechnic University, Hong Kong SAR (email: weiren\_thu@163.com and jiannong.cao@polyu.edu.hk).}
\thanks{Huansheng Yi, Jianqi Wang, and Fugui Qi are with the Department of Biomedical Engineering, Fourth Military Medical University, and Innovation Research Institute, Xijing Hospital, Fourth Military  Medical University, Xi'an, China, 710032, China (email: Yihsbme@outlook.com, wangjq@fmmu.edu.cn, and Qifgbme@outlook.com).}
\thanks{Kaiyue Hou is with the Department of Electrical and Electronic Engineering, The Hong Kong Polytechnic University, Hong Kong SAR (email: kaiyue.hou@connect.polyu.hk).} 
\thanks{Miaoyang Hu is with the Department of Cardiology, Xijing Hospital, Fourth Military Medical University, Xi’an, 710032, China (email: humiaoyang@qq.com).}
}

\markboth{Wei Ren \MakeLowercase{\textit{et al.}}: Noncontact Multi-Point Vital Sign Monitoring with mmWave MIMO Radar}%
{Wei Ren \MakeLowercase{\textit{et al.}}: Noncontact Multi-Point Vital Sign Monitoring with mmWave MIMO Radar}


\maketitle

\begin{abstract}
Multi-point vital sign monitoring is essential for providing detailed insights into physiological changes. Traditional single-sensor approaches are inadequate for capturing multi-point vibrations. 
Existing contact-based solutions, while addressing this need, can cause discomfort and skin allergies, whereas noncontact optical and acoustic methods are highly susceptible to light interference and environmental noise. 
\revised{
In this paper, we aim to develop a non-contact, multi-point vital sign monitoring technique using MIMO radar, focused on physically differentiating and precisely measuring chest-wall surface vibrations at multiple points induced by cardiopulmonary mechanical activity.
} 
The primary challenges in developing such a technique involve developing algorithms to extract and separate entangled signals, as well as establishing a reliable method for validating detection accuracy.
To address these limitations, we introduce MultiVital, a wireless system that leverages mmWave Multiple-input Multiple-output (MIMO) radar for synchronous multi-point vital sign monitoring. 
\revised{It integrates two reference modalities: five-channel seismocardiography (SCG) sensors and a one-channel electrocardiogram (ECG) electrode, enabling comprehensive radar-based research and performance validation across multiple physiological metrics.
}
Additionally, we have developed a multi-modal signal processing framework, consisting of a radar signal processing module, an SCG calibration module, and a spatial alignment scheme.
To evaluate the radar signal processing module, we conducted mathematical derivation and simulation. 
The experimental results indicate that the noncontact MultiVital system achieves multi-point synchronous monitoring with high precision, highly consistent with the results from reference modalities.
This system enables the precise detection of subtle cardiopulmonary movements in different regions of the human body, providing more accurate and comprehensive information for cardiopulmonary health monitoring.
\end{abstract}


\section{Introduction}
\label{sec:introduction}
\IEEEPARstart{V}{ital} \revised{sign detection by measuring chest-wall surface vibrations induced by cardiopulmonary mechanical activity is critical for monitoring cardiovascular and respiratory health, as well as diagnosing related diseases \cite{zakeri2016analyzing,zhao2018noncontact,wen2023noncontact}.}
More specifically, surface vibrations generated by different chambers of the heart and lungs exhibit inherent positional specificity, even within the same cardiac or respiratory cycle \cite{crow1994relationship}, providing more localized and detailed information about physiological changes. Previous single-sensor-based solutions are significantly limited, as they only capture the aggregate vibrations from the entire sternal surface, lacking the positional resolution needed for detailed analysis\cite{gu2013hybrid,gu2012accurate,wang2014application, ren2016phase,fathy2017overview,xia2020delineation,rong2019remote,dai2021enhancement,li2021through}.

The multi-point vibration synchronous measurement solution enables position-specific sensing. 
Multiple contact or non-contact sensors array placed on/over the chest provide a wealth of diversified information to analyze complex heartbeat and breath motion dynamics with better spatial and temporal resolution, which facilitates detecting more location-specific motion of each valve or chamber and identifying the signal source. 
Presently, multi-point measurements are mainly based on contact and semi-contact methods. 
Since Okada  first proposed the multi-channel recording of heart sounds, many sensing technologies were applied to this \cite{okada1982chest}, like the microphone array \cite{cozic1998development}, fiber optic technology \cite{lo2021multi}, and infrared (IR) cameras and optical markers \cite{shafiq2014surface,shafiq2017multimodal}.
Additionally, the hottest study is multichannel seismocardiography (mchSCG or MSCG) to understand the distribution of vibration waves on the chest wall by placing multiple accelerometers on the chest \cite{lin2016identification,lin2018realization,munck2020multichannel}. 
Based on the MSCG, rich location-specific feature points in a cardiac cycle corresponding to the multiple valvular auscultation locations were identified \cite{lin2016identification} combined with synchronous electrocardiogram (ECG) and echocardiography recording.
Unfortunately, the above-mentioned contact solutions would inevitably result in unpleasant user experience and even skin allergies, while sensors or electrodes are attached to the skin for a long time. 
On the other hand, the non-contact solutions based on optics or acoustics are highly susceptible to ambient light or noise, rendering them unsuitable for clinical diagnosis and home health monitoring \cite{van2020camera,shirkovskiy2018airborne,sadhukhan2023analysis,struijk2016heart}.

\revised{In contrast to wearable devices and cameras, wireless devices such as mmWave Multiple-input Multiple-output (MIMO) radar present a more promising approach for monitoring multi-point vital sign signals \cite{li2022remote,eder2023sparsity,bauder2024mm,wang2025contactless}.} 
mmWave MIMO radar systems utilize high frequency signals in the millimeter wave spectrum (typically 30 GHz to 300 GHz) and the MIMO architecture (multiple transmitting and receiving antennas) to achieve high resolution detection \cite{dai2021enhancement,li2023motion,chen2022contactless}, thus separating multiple scattering points from human chest. 

These mmWave MIMO radars can be conveniently installed on walls and ceilings of both homes and hospitals, ensuring they integrate smoothly into people's daily routines without causing any disturbances. 
The radars work by emitting electromagnetic waves into the space, then collecting the waves reflected back by individuals. 
\revised{Advanced signal processing techniques are employed to analyze these reflections, allowing for an accurate assessment of a person's condition \cite{ren2021vital, qiao2022contactless, zhang2019bioradar,xia2020delineation, marnach2024comparison,sameera2024respiratory}. }
By using mmWave radar instead of wearable devices or cameras, individuals no longer need to experience discomfort caused by wearable devices or have the privacy concerns introduced by cameras. 
However, the development of such a wireless solution using mmWave MIMO radar is nontrivial. 
Developing feasible radar signal processing procedure for synchronous monitoring of multi-point vital signs and building a hardware system to validate the validity and effectiveness of the radar-based solution are keys to the successful implementation of this radar-based wireless solution.

In this study, we introduce MultiVital, a wireless solution for multi-point vital sign monitoring using mmWave MIMO radar. 
This comprehensive solution encompasses both a hardware system and algorithm development. 
The contributions of our radar-based solution for multi-point vital sign monitoring include
\begin{itemize}
    \item Building a new hardware system, namely MultiVital, which integrates a mmWave MIMO radar system, five-channel SCG sensors, and one-channel ECG electrodes. This hardware system facilitates synchronous monitoring of multi-point vital signs and provides two reference modalities for evaluating the performance of radar system.  
    \item Designing the overall signal processing framework to \revised{physically differentiate and accurately monitor five-point vital signs}. This framework consists of a radar signal processing module, an SCG calibraion module, and a scheme for their spatial alignment. 
    \item \revised{Conducting theoretical derivation, simulation and experiments to validate the feasibility and effectiveness of signal processing module to extract multi-point vital signs.}  
\end{itemize}   
    
The remaining sections of this paper are organized as follows. 
In Section~\ref{sec:radar signal processing}, the radar signal processing steps are presented. 
In Section~\ref{sec:MultiVital system}, a detailed description of the MultiVital hardware system and the overall signal processing framework of the MultiVital system are presented. 
In Section~\ref{sec:simulation} and Section~\ref{sec:experiment}, the simulation and experimental results are demonstrated and discussed, respectively. 
In Section~\ref{sec:conclusion}, the conclusion is drawn.

\section{Radar Signal Processing}
\label{sec:radar signal processing}

\begin{figure}[htbp]
\centerline{\includegraphics[scale=0.5]{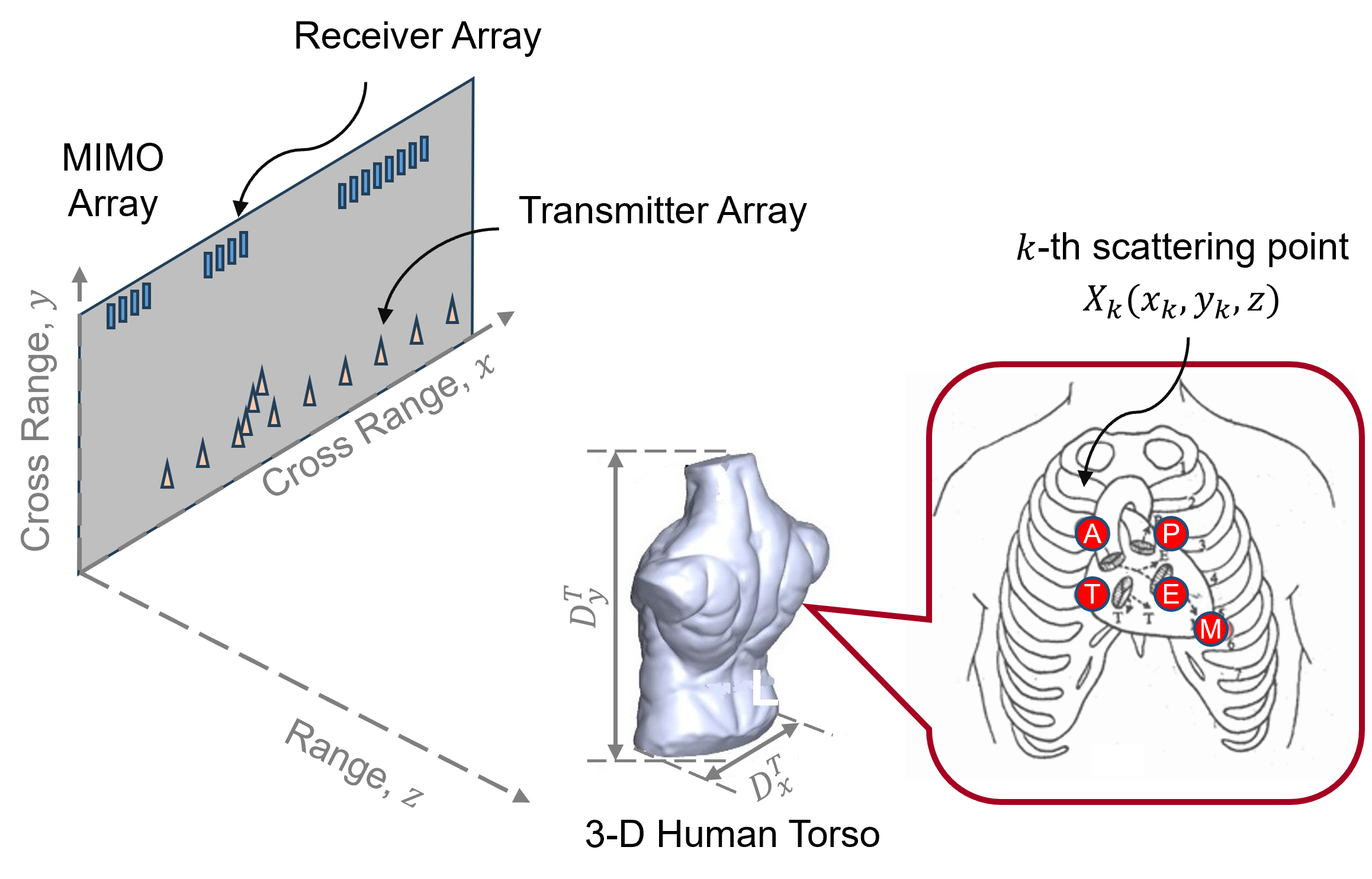}}
\caption{\revised{Research scenario of multi-point vital sign monitoring with mmWave MIMO radar. The MIMO radar system is placed in front of the human chest wall and used to estimate the motion of the five scattering points of interest, i.e., point A, P, T, E, and M.
}}
\vspace{-5mm}
\label{fig:spatial alignment scheme}
\end{figure}

Fig.~\ref{fig:spatial alignment scheme} illustrates the research scenario, where the radar system is placed in front of the human chest wall. The task is to estimate the motion (displacement versus time) of the five scattering points of interest (point A, P, T, E, and M) on the human chest wall using radar system.  
In this section, we derive the signal processing steps to estimate multi-point displacements of the human chest wall    from the collected radar signals.

Here, the radar signals transmitted by the mmWave MIMO radar are Frequency Modulated Continuous Wave (FMCW) signals. 
The transmitted FMCW signal with a linear frequency modulation pattern can be expressed as 
\begin{equation}
s(t) = A_{t}{\cos\left( {2\pi f_{c}t + \pi\frac{B}{T}t^{2}} \right)},
\end{equation}
where $A_t$ is the transmitted signal amplitude, $f_c$ is carrier frequency, $B$ is the bandwidth, and $T$ is the pulse duration.

Here, we consider a scenario of a target consisting of $K$ scattering points in total and the $k$-th scattering point at the distance $R_k (t)$, elevation angle $\theta_k$ and azimuth angle $\phi_k$. 
The steering vectors for transmitting antenna array and receiving antenna array are $\mathbf{a}\left( {\theta_{k},\phi_{k}} \right),~\mathbf{a} \in \mathbb{C}^{N_{\text{Tx}} \times 1}$ and $
\mathbf{b}\left( {\theta_{k},\phi_{k}} \right),~\mathbf{b} \in \mathbb{C}^{N_{\text{Rx}} \times 1}$, respectively, which are
\begin{equation} \label{eq:transmitting vector a}
\mathbf{a}\left( {\theta_{k},\phi_{k}} \right) = \left\lbrack {a_{1}\left( {\theta_{k},\phi_{k}} \right),~\ldots,~a_{N_{\text{Tx}}}\left( {\theta_{k},\phi_{k}} \right)} \right\rbrack^{T},
\end{equation}
and
\begin{equation} \label{eq:receiving vector b}
    \mathbf{b}\left( {\theta_{k},\phi_{k}} \right) = \left\lbrack {b_{1}\left( {\theta_{k},\phi_{k}} \right),~\ldots,~b_{N_{\text{Rx}}}\left( {\theta_{k},\phi_{k}} \right)} \right\rbrack^{T}.
\end{equation}
The received radar signals of all $\left( {N_{\text{Tx}} \cdot N_{\text{Rx}}} \right)$ channels are
\begin{equation}
    \mathbf{X}_{k} = \mathbf{b}\left( {\theta_{k},\phi_{k}} \right)\mathbf{a}^{H}\left( {\theta_{k},\phi_{k}} \right)s\left( {t - \tau_{k}(t)} \right),
\end{equation}
where
\begin{equation}
\tau_{k}(t) = 2R_{k}(t)/c
\end{equation}
is the time delay for $k$-th scattering point, and $c$ is the speed of light. The received radar signal of the channel from $n_t$-th transmitting antenna and $n_r$-th receiving antenna is
\begin{equation}
\begin{split}
\!\!\!\!&\!\!s_{r}\left( {n_{r},n_{t},k,t} \right) \\
\!\!\!\!\!&\!\!= b_{n_{r}}~\left( {\theta_{k},\phi_{k}} \right)a_{n_{t}}^{*}\left( {\theta_{k},\phi_{k}} \right)As\left( {t - \tau_{k}(t)} \right) \\
\!\!\!\!\!&\!\!= C\left( {n_{r},n_{t},k} \right){\cos\!\left( {2\pi f_{c}\left( {t - \tau_{k}(t)} \right)\! +\! \pi\frac{B}{T}\left( {t - \tau_{k}(t)} \right)^{2}}\! \right)},
\end{split}
\end{equation}
where $A$ is the amplitude coefficient $A=A_r/A_t$ and $C\left( {n_{r},n_{t},k} \right) = b_{n_{r}}~\left( {\theta_{k},\phi_{k}} \right)a_{n_{t}}^{*}\left( {\theta_{k},\phi_{k}} \right)A$. $A_r$ is the amplitude of the received signal. $(\cdot)^*$ denotes conjugate value.

Then the received signal is multiplied by the reference signal, and then passed through a low-pass filter to obtain the intermediate frequency (IF) signal of an in-phase channel as
\begin{equation}
\begin{split}
\!\!\!&\!\!\!\!s_{if}^{(\text{I})}\left( {n_{r},n_{t},k,t} \right) \\
\!\!&\!\!\!\!\!\!= f_{LPF}\left\{ {s(t)s_{r}\left( {n_{r},n_{t},k,t} \right)} \right\} \\
\!&\!\!\!\!\!\!=\!\! \frac{1}{2}A_{t}C\!\left( \!{n_{r},n_{t},k}\! \right){\!\cos\!\left(\! {2\pi\frac{B}{T}\tau_{k}(t)t \!+\! 2\pi\! f_{c}\tau_{k}(t) \!- \!\pi\frac{B}{T}\tau_{k}^{2}(t)} \!\!\right)}.
\end{split}
\end{equation}
The corresponding complex signal is (ignoring constant amplitude $\frac{1}{2}A_t$ for simplicity)
\begin{equation} \label{eq:s_if}
\begin{split}
\!\!\!\!\!\!\!&s_{if}\left( {n_{r},n_{t},k,t} \right) \\
\!\!\!\!\!\!\!&= \!C\!\left( {n_{r},n_{t},k} \right){\exp{j\!\!\left( {2\pi\frac{B}{T}\tau_{k}(t)t\!+ \!2\pi f_{c}\tau_{k}(t)\! - \!\pi\frac{B}{T}\tau_{k}^{2}(t)}\! \right)}} \\
\!\!\!\!\!\!\!&\approx C\left( {n_{r},n_{t},k} \right){\exp{j\left( {2\pi\frac{B}{T}\tau_{k}(t)t + 2\pi f_{c}\tau_{k}(t)} \right)}}.
\end{split}
\end{equation}
In (\ref{eq:s_if}), the last term of the second formula ($
- \pi\frac{B}{T}\tau_{k}^{2}(t)$) can be ignored \cite{ren2023tracking}. 

Here suppose that a total of $M$ frames of signals are collected, and the signals of $m$-th frame in time and frequency domain are $
s_{if}\left( {n_{r},n_{t},k,t,m} \right)$ in (\ref{eq:s_if_with_m}) and $S_{if}\left( {n_{r},n_{t},k,f,m} \right)$ in (\ref{eq:S_if_with_m}), respectively.  
$S_{if} (n_r,n_t,k,f,m)$ is the Fourier transform (FT) of $s_{if}\left( {n_{r},n_{t},k,t,m} \right)$.  
\begin{equation} \label{eq:s_if_with_m}
\begin{split}
&s_{if}\left( {n_{r},n_{t},k,t,m} \right) \\
&= C\left( {n_{r},n_{t},k} \right){\cos\left( {2\pi\frac{B}{T}\tau_{k}(m)t + 2\pi f_{c}\tau_{k}(m)} \right)}
\end{split}    
\end{equation}
\begin{equation}\label{eq:S_if_with_m}
\begin{split}
&\!\!\!\! S_{if}\left( {n_{r},n_{t},k,f,m} \right) \\
&\!\!\!\!\!= {\int_{- T/2}^{T/2}{s_{if}\left( {n_{r},n_{t},k,t,m} \right)e^{- j2\pi ft}dt}} \\
&\!\!\!\!\! =\! C\left( {n_{r},n_{t},k} \right)\!\frac{e^{j2\pi f_{c}\tau_{k}{(m)}}}{\pi\!\left( {\frac{B}{T}\tau_{k}(m) - f} \right)}{\sin\!\!\left( \!{\pi T\left(\! {\frac{B}{T}\tau_{k}(m) \!-\! f}\! \right)}\! \right)} \\
&\!\!\!\!\! = C\left( {n_{r},n_{t},k} \right)e^{j2\pi f_{c}\tau_{k}{(m)}}T\mathrm{sinc}\!\left( \!{T\left(\! {\frac{B}{T}\tau_{k}(m) \!- \!f}\! \right)}\! \right),
\end{split}
\end{equation}
where $\tau_k(m)$ is the time delay of $m$-th frame. For $m$-th frame, the maximum modulus points $
S_{if}\left( {n_{r},n_{t},k,f_{0}(m),m} \right)$ can be obtained by setting $
f_{0}(m) = \frac{B}{T}\tau_{k}(m)$. 
The maximum modulus points of $M$ frames are represented as $
\mathbf{z}$ (ignoring the amplitude $
T\mathrm{sinc}\left( {T\left( {\frac{B}{T}\tau_{k}(m) - f_{0}(m)} \right)} \right)$)
\begin{equation}
\begin{split}
&\!\!\!\!\! \mathbf{z}\left( {n_{r},n_{t},k} \right) \\
& \!\!\!\!\!\!\!= \!\!\Bigl\lbrack {S_{if}\!\left( {n_{r},n_{t},k,f_{0}(1),1} \right)\!,S_{if}\left( {n_{r},n_{t},k,f_{0}(2),2} \right),\ldots, }\Bigr. \\
& \qquad \qquad \qquad \qquad \Bigl. {S_{if}\left( {n_{r},n_{t},k,f_{0}(M),M} \right)} \Bigr\rbrack \\
& \!\!\!\!\!\!\!\approx\! C\!\left( {n_{r},n_{t},k} \right)\!\left\lbrack {e^{j2\pi f_{c}\tau_{k}{(1)}},e^{j2\pi f_{c}\tau_{k}{(2)}},\ldots,e^{j2\pi f_{c}\tau_{k}{(M)}}}\! \right\rbrack,
\end{split}
\end{equation}
and their phases are
\begin{equation}
\begin{split}
& \!\!\!\!\mathbf{g}\left( {n_{r},n_{t},k} \right) \\
&\!\!\!\!\! = {\arg\left( {\mathbf{z}\left( {n_{r},n_{t},k} \right)} \right)} \\
&\!\!\!\!\! = {{2\pi f_{c}\left\lbrack {\tau_{k}(1),\tau_{k}(2),\ldots,\tau_{k}(M)} \right\rbrack + \mathit{\arg}}\left( {C\left( {n_{r},n_{t},k} \right)} \right)} \\
&\!\!\!\!\! = {{\frac{4\pi}{\lambda}\left\lbrack {R_{k}(1),R_{k}(2),\ldots,R_{k}(M)} \right\rbrack + \mathit{\arg}}\left( {C\left( {n_{r},n_{t},k} \right)} \right)},
\end{split}
\end{equation}
where $\arg( \cdot )$ represents the function of computing the phase, and $
\lambda = c/f_{c}$ is the wavelength. 
From $
\mathbf{g}\left( {n_{r},n_{t},k} \right)$, we can estimate the displacement $
\left\lbrack {R_{k}(m)} \right\rbrack_{m = 1,\ldots,M}$ of the $k$-th scattering point.
In the above derivation, we focus on the signals from a single scattering point. The received signals of all $K$ scattering points can be further expressed as 
\begin{equation}
s_{r}\left( {n_{r},n_{t},t} \right) = {\sum_{k = 1}^{K}{s_{r}\left( {n_{r},n_{t},k,t} \right)}},
\end{equation}
and the corresponding final phase signal is
\begin{equation}
\mathbf{g}\left( {n_{r},n_{t}} \right) = {\sum_{k = 1}^{K}{\mathbf{g}\left( {n_{r},n_{t},k} \right)}}.
\end{equation}
The displacement $\left\lbrack {R_{k}(m)} \right\rbrack_{m = 1,\ldots,M,k = 1,\ldots,K}$ of each scattering point is estimated by involving Direction-of-arrival (DOA) estimation method, 
\revised{such as Fast Fourier Transform (FFT).}

\section{MultiVital System and Overall Processing Procedure}
\label{sec:MultiVital system}
In this section, we first introduce the MultiVital hardware system that integrates a mmWave MIMO radar system, five-channel SCG sensors and one-channel ECG electrodes. After that, we present the overall signal processing framework.  

\subsection{MultiVital Hardware System}

\begin{figure*}[htbp]
\centerline{\includegraphics[scale=0.45]{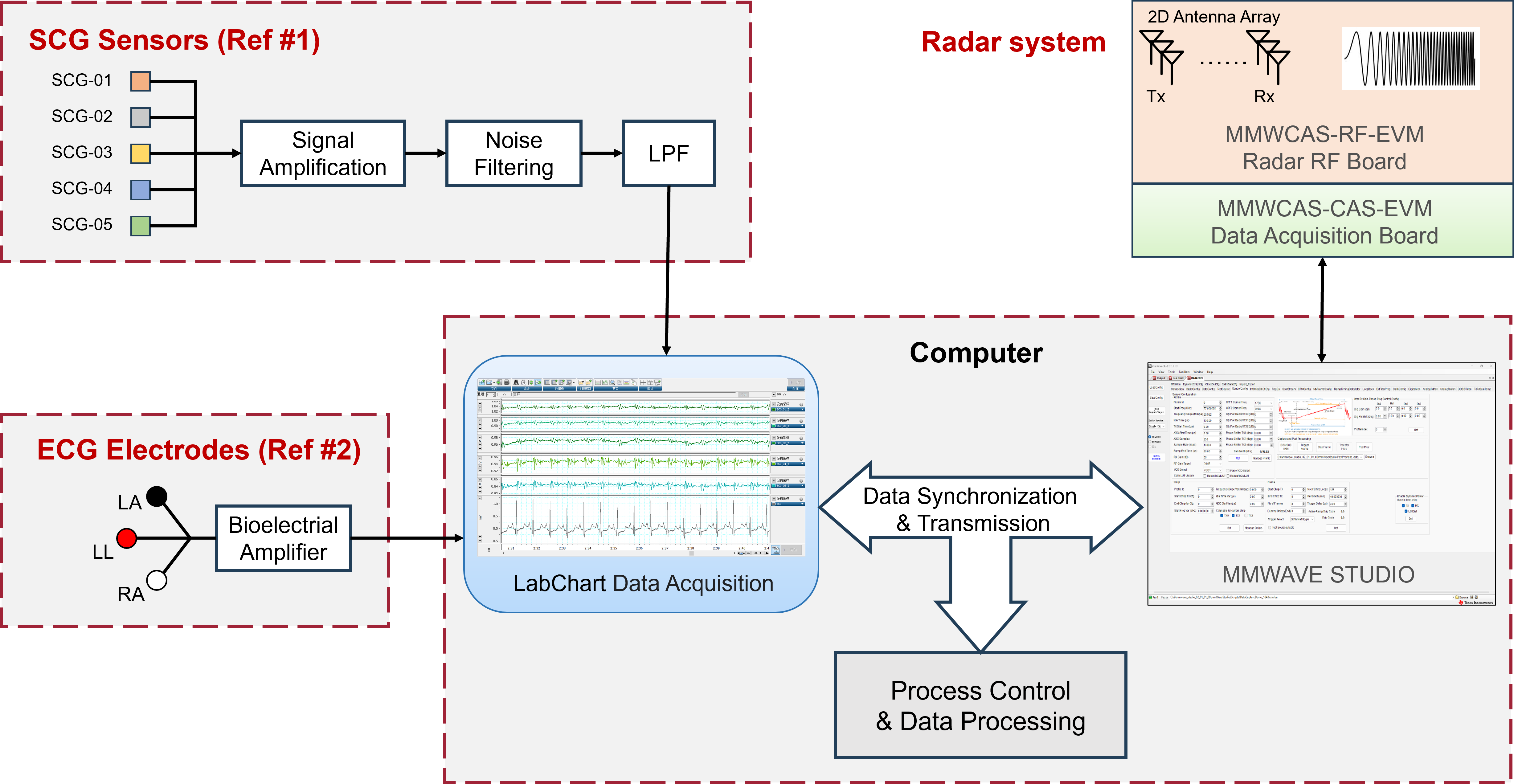}}
\caption{\revised{
The MultiVital system integrates an mmWave MIMO radar system along with two reference systems including five-channel SCG sensors and one-channel ECG electrodes.}}
\label{fig:MultiVital system}
\end{figure*}

\begin{figure}[htbp]
\centerline{\includegraphics[scale=0.35]{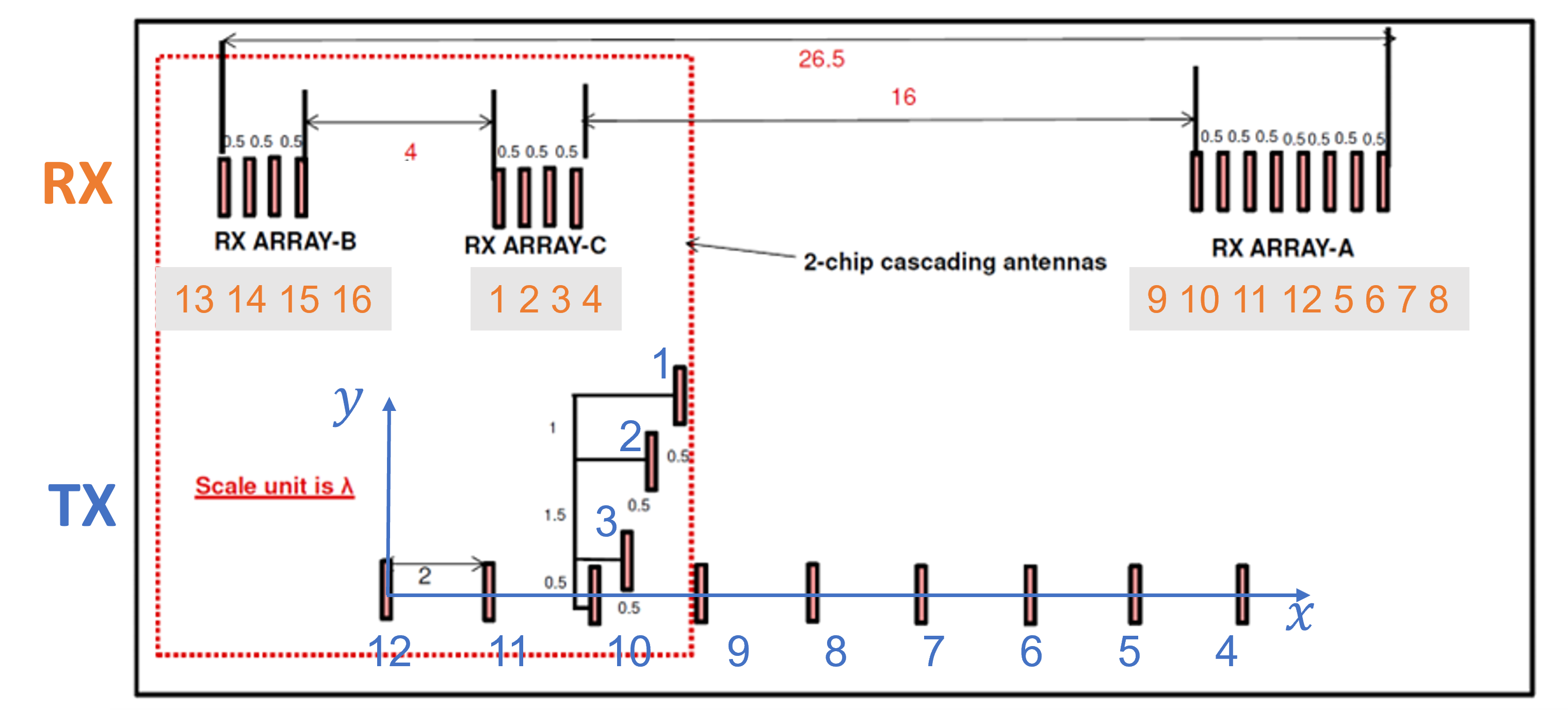}}
\caption{Antenna array positions \cite{TI_MMWCAS_RF_EVM}.}
\label{fig:antenna array positions}
\vspace{-5mm}
\end{figure}

\begin{figure}[htbp]
\centerline{\includegraphics[scale=0.40]{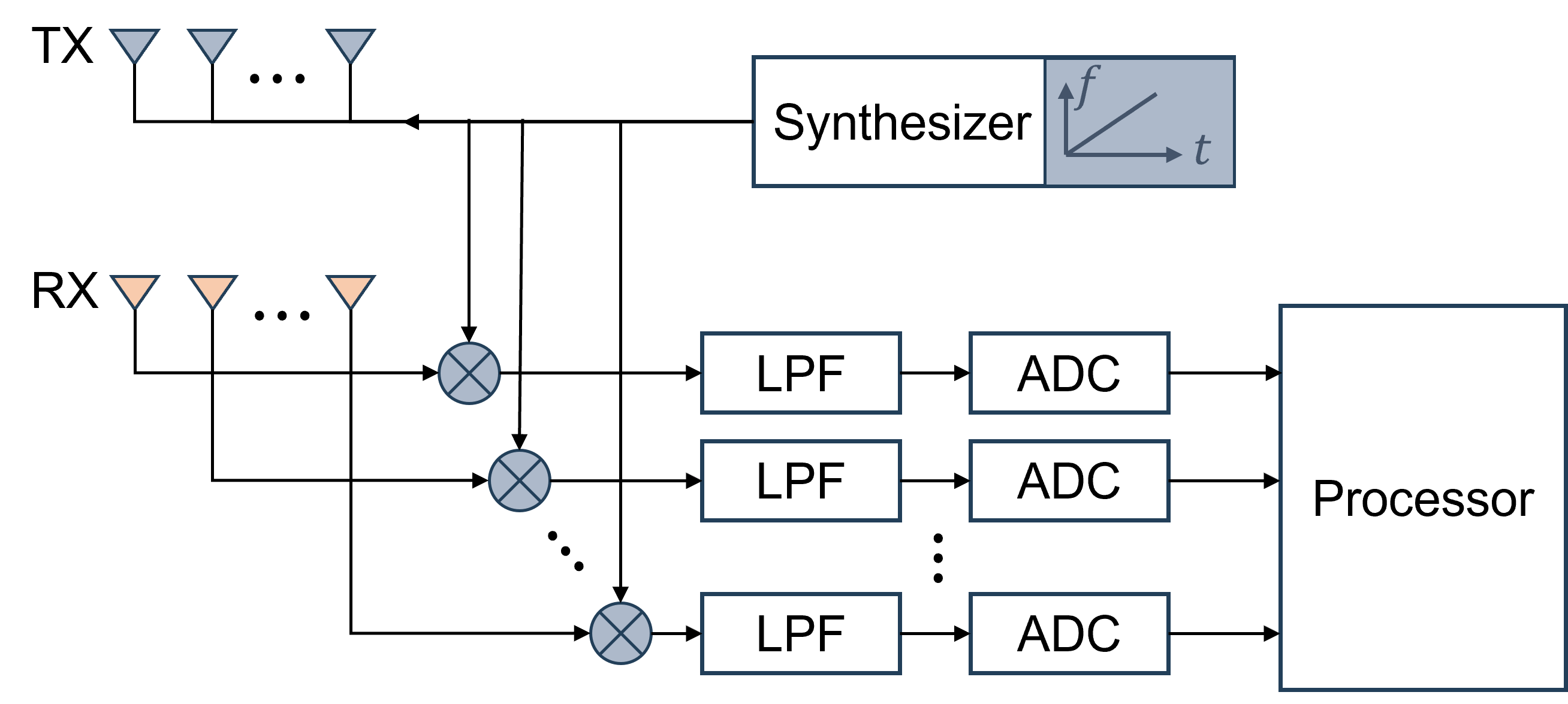}}
\caption{Radar system diagram.}
\label{fig:radar system diagram}
\vspace{-5mm}
\end{figure}

\begin{figure*}[htbp]
\centering
\begin{minipage}[t]{0.8\linewidth}
\centering
\includegraphics[width=\linewidth]{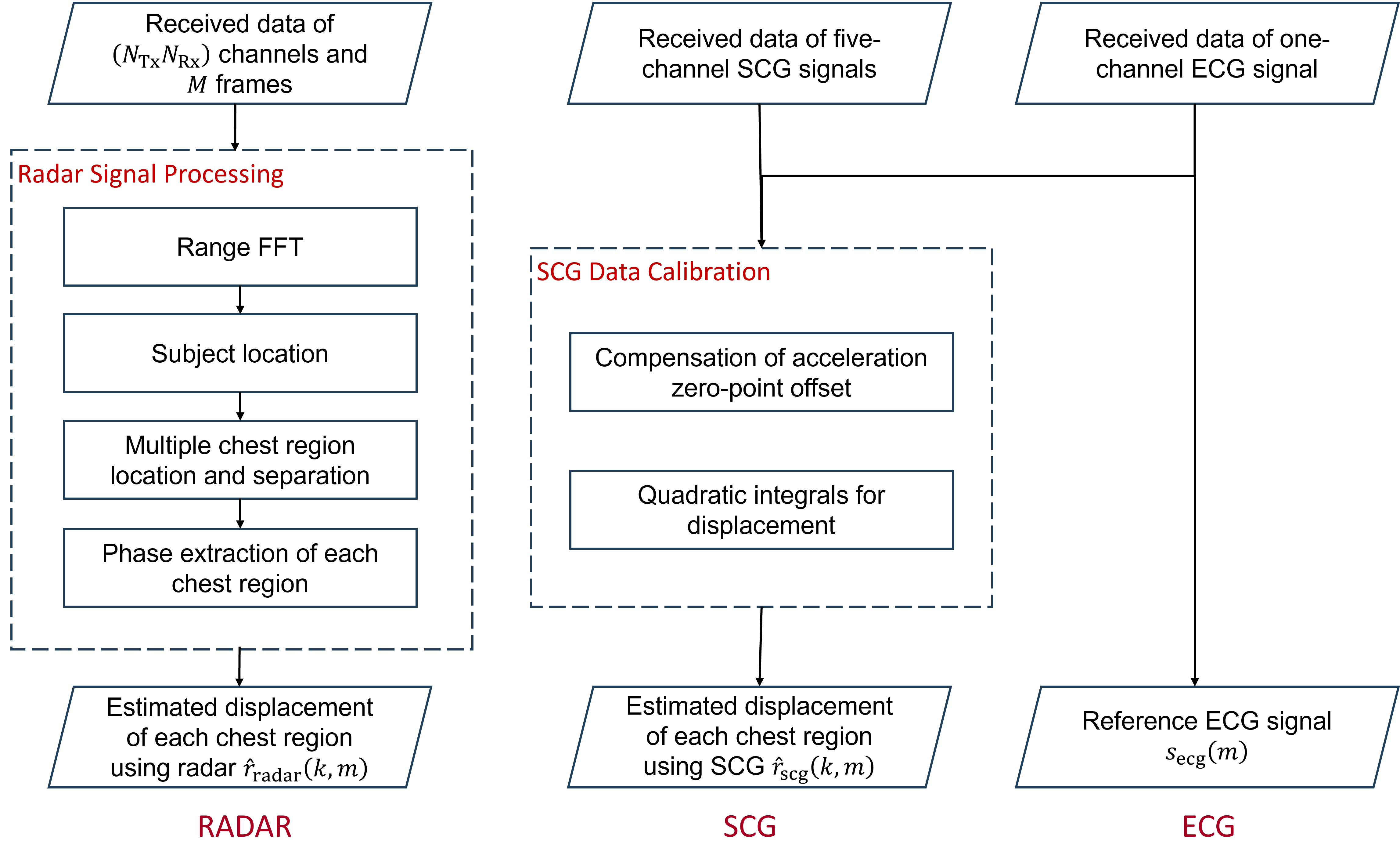}
\caption{The overall signal processing framework of MultiVital system.}
\label{fig:overall processing framework}
\end{minipage}
\begin{minipage}[t]{0.85\linewidth}
\centering
\vspace{10mm}
\includegraphics[width=\linewidth]{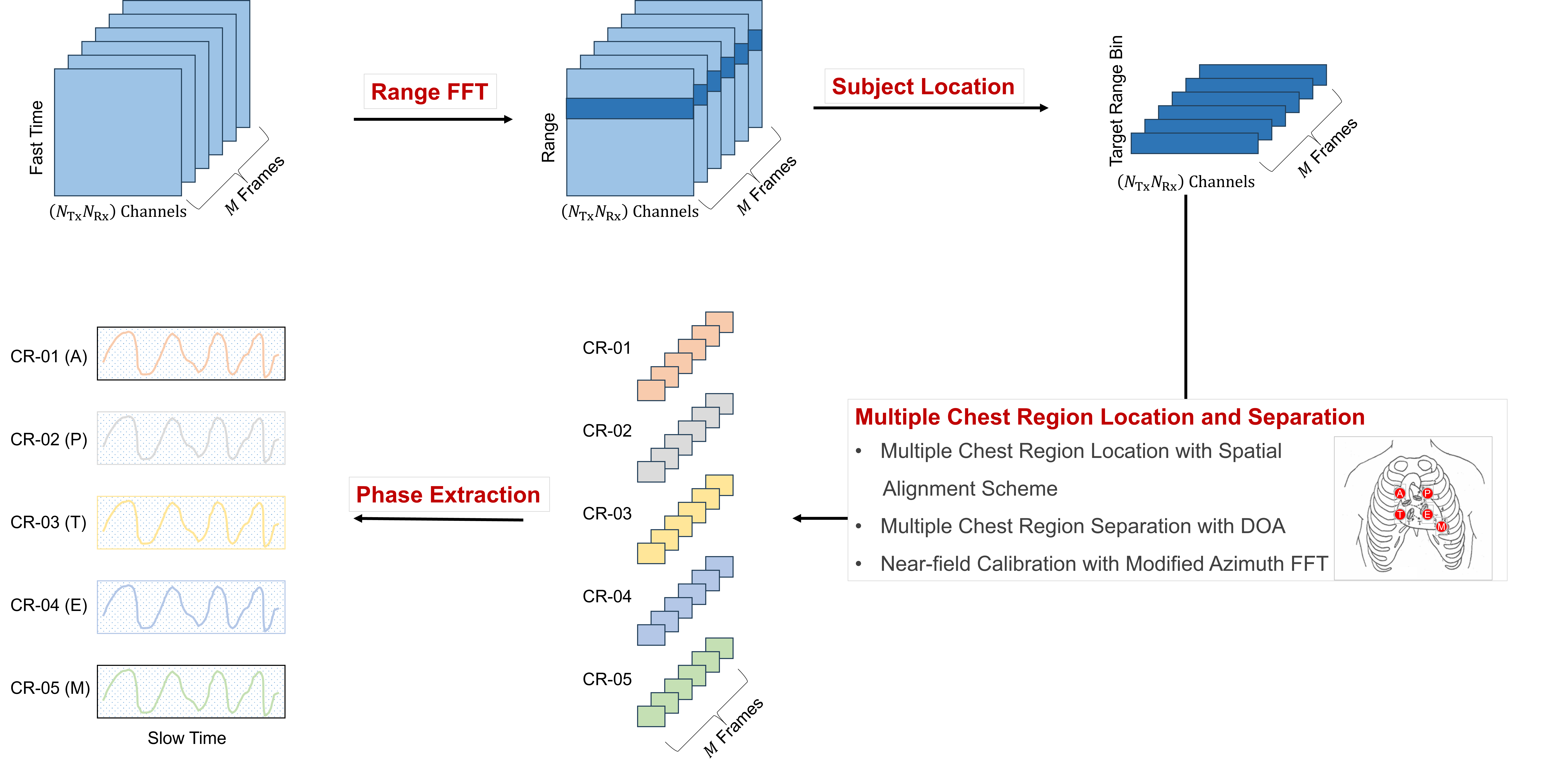}
\caption{\revised{Radar signal processing module. CR-$i$ represents the $i$-th chest region.}}
\label{fig:radar signal processing module}
\end{minipage}
\begin{minipage}[t]{0.8\linewidth}
\centering
\vspace{10mm}
\includegraphics[width=\linewidth]{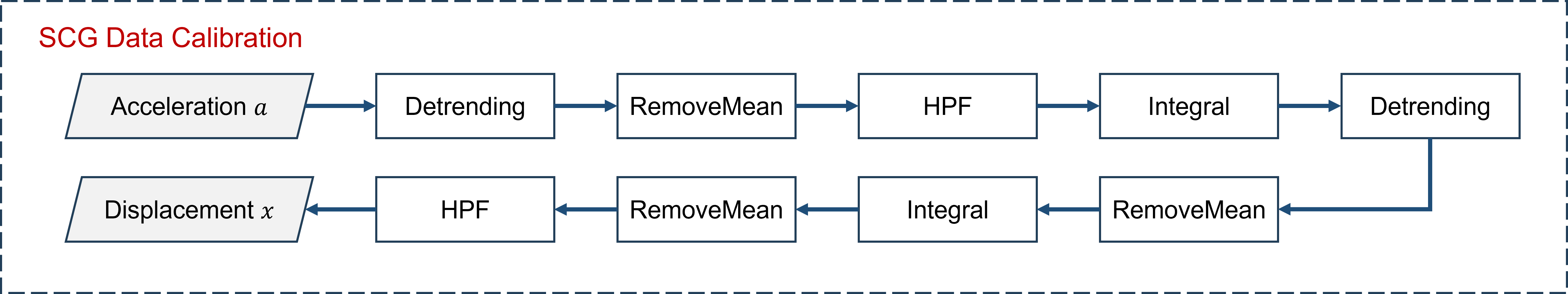}
\caption{SCG calibration module. Detrending is to remove the best straight-fit line from the data and returns the remaining data; RemoveMean denotes the operation of removing the mean value from the data; HPF denotes high pass filtering.}
\label{fig:SCG calibration module}
\end{minipage}
\end{figure*}




Fig.~\ref{fig:MultiVital system} shows the MultiVital hardware system which integrates a mmWave MIMO radar system, five-channel SCG sensors and one-channel ECG electrodes.

\revised{In this work, the system uses MMWCAS-RF-EVM and MMWCAS-DSP-EVM for mmWave radar signal transmission and data acquisition \cite{TI_MMWCAS_RF_EVM,MMWCAS-DSP-EVM}. }
The MMWCAS-RF-EVM primarily consists of a cascaded mmWave radar transceiver module formed by four AWR2243 chips. 
It supports up to 12 transmit (Tx) and 16 receive (Rx) antennas, totaling 192 virtual receive channels. 
The antenna array and radar system diagram are shown in Fig.~\ref{fig:antenna array positions} and Fig.~\ref{fig:radar system diagram}, respectively.
\revised{
This setup forms a MIMO virtual antenna array, consisting of a uniform linear array (ULA) with 86 elements and a spacing of $\lambda/2$ in the azimuth direction, along with a minimum redundancy array (MRA) configuration in the elevation direction.
}
The radar is capable of operating in the 76~GHz to 81~GHz frequency band for FMCW transmission and reception, with a distance resolution of up to 3.75~cm, an antenna field of view of approximately $
\pm 70{^\circ}$, angular resolutions of about $1.4^{\circ}$ in azimuth and $18^{\circ}$ in elevation.
\revised{The PowerLab 16/35 DAQ is used for data acquisition, providing up to 16 input channels, a 400 ksps sampling rate, and multi-host connectivity \cite{DAQ}. The DAQ's key features include high fidelity signal reproduction, multiple input interfaces, and customized waveform output.}


To synchronously and accurately acquire ECG signals, the system also incorporates Dual Bio Amp (from AD Instruments) as the electrocardiography acquisition and amplification system \revised{\cite{Bio_Amps}}. 
This device can capture bioelectrical signals ranging from $\pm 5~\mu\mathrm{V}$ to $\pm 100~\mathrm{mV}$. 
It features a low-noise, high-gain differential amplifier specifically designed for bio-signal measurement, as well as built-in software-controlled low-pass, high-pass, and notch filters to eliminate unwanted signal frequencies for specific applications.
Each channel of the acquisition card connects to an ADXL354 (from ANALOG DEVICES) micro-electromechanical system (MEMS) accelerometer, which boasts a compact size and high integration level \revised{\cite{AnalogDevices_ADXL354}}. 
This facilitates the stable and secure attachment of the sensor to the chest surface. 
The sensor supports triaxial acceleration measurement ranges of $\pm 2~\mathrm{g}$ and $\pm 4~\mathrm{g}$.

LabChart software, produced by AD Instruments, is used for data recording and preliminary processing on the host computer \cite{ADInstruments_LabChart}. 
LabChart is an integrated software solution for data acquisition, recording, and analysis, requiring only a USB connection to interface with the PowerLab acquisition system. 
It allows for configuration settings such as sampling rate, voltage range, and channel mapping, significantly simplifying the use of data acquisition devices, facilitating quick parameter setting, and enabling the commencement of experiments.

\subsection{Overall Signal Processing Framework}
\label{subsec:overall processing framework}
The overall signal processing framework is shown in Fig.~\ref{fig:overall processing framework}.
The inputs of the processing framework are radar data (received data of ($N_{\text{Tx}} N_{\text{Rx}}$) channels and $M$ frames), SCG data (received data of five-channel SCG signals), and ECG data (received data of one-channel ECG data). 
The expected outputs are radar estimated data (estimated displacement of each chest region using radar, ${\hat{r}}_{\text{radar}}\left( {k,m} \right)$, SCG estimated data (estimated displacement of each chest region using SCG, ${\hat{r}}_{\text{scg}}\left( {k,m} \right)$), and reference ECG signal, $s_{\text{ecg}}(m)$. 

The ECG signals can be used directly without any further processing. 
To estimate the displacement of multiple chest regions using radar and SCG sensors, we designed radar signal processing module and SCG signal calibration module, respectively.

1) Radar signal processing module.
The radar signal processing module is shown in Fig.~\ref{fig:radar signal processing module}, consisting of four steps as follows.
\begin{itemize}
    \item Range fast Fourier transform (FFT). The first step is to perform range FFT along fast-time axis to get the range profile $S_{if}\left( {n_{r},n_{t},f,m} \right)$.
    \item Subject location. The subject is located by finding the maximum of the range profiles of all ($N_{\text{Tx}}N_{\text{Rx}}$) channels and all $M$ frames. Then we take the signals of the range bin where the subject is located as the signals of interest $S_{if}\left( {n_{r},n_{t},f_{0},m} \right)$.
    \item \revised{Multiple chest region location and separation. This is accomplished by incorporating three major components: a spatial alignment scheme to locate multiple chest regions, a DOA technique to separate these regions, and a near-field calibration with modified azimuth FFT to enhance accuracy. Together, these techniques enhance the estimation accuracy of the five chest regions of interest, denoted as $S_{if}\left( {k,f_{0},m} \right),~k = 1,2,\ldots,5$.}
    \item Phase extraction of each chest region. Then the phases of the signals $S_{if}\left( {k,f_{0},m} \right)$ are extracted as the displacements of the five interested chest regions, i.e., $g\left( {k,f_{0},m} \right) = {\arg\left( {S_{if}\left( {k,f_{0},m} \right)} \right)},~k = 1,2,\ldots,5$.
\end{itemize}

2) SCG signal calibration module.
The SCG signal calibration module is to first compensate the acceleration zero-point offset and then calculate the displacements of the five focused chest regions. 
Fig.~\ref{fig:SCG calibration module} presents the detailed SCG module, which consists of a series of operations like detrending, removing mean value, high pass filtering, and integral operation. 
By these steps, we can estimate the accurate displacements of five regions of interest and use them as the ground truth to assess the performance of radar signal processing procedure.

\vspace{-2mm}\section{Simulation and Discussion}
\label{sec:simulation}

\begin{table}[]
\centering
\caption{mmWave MIMO radar parameters for simulation.}
\label{table:radar parameters}
\begin{tabular}{l|l|l}
\toprule
\textbf{Parameter Name}       & \textbf{Symbol}   & \textbf{Value} \\
\midrule
Carrier Frequency             & $f_c$              & $77~\mathrm{GHz}$         \\
Pulse Repetition Time         & PRT               & $85.3~\mu\mathrm{s}$        \\
Frame Duration                & $T_{\text{Frame}}$          & $0.135~\mathrm{s}$        \\
Number of ADC Samples         & $N_{\text{ADCSamples}}$     & $512$            \\
Sampling   Frequency          & $f_s$              & 7 MHz          \\
Chirp Rate                    & $K_{\text{Chirp}}$          & $63.005~\mathrm{MHz}/\mu\mathrm{s}$  \\
Number of Chirps in One Frame & $N_{\text{ChirpsOneFrame}}$ & $128$            \\
Total Frames                  & $N_{\text{Frame}}$          & $50$             \\
\bottomrule
\end{tabular}
\vspace{-1mm}\end{table}

\begin{figure}[htbp]
\centerline{\includegraphics[scale=0.5]{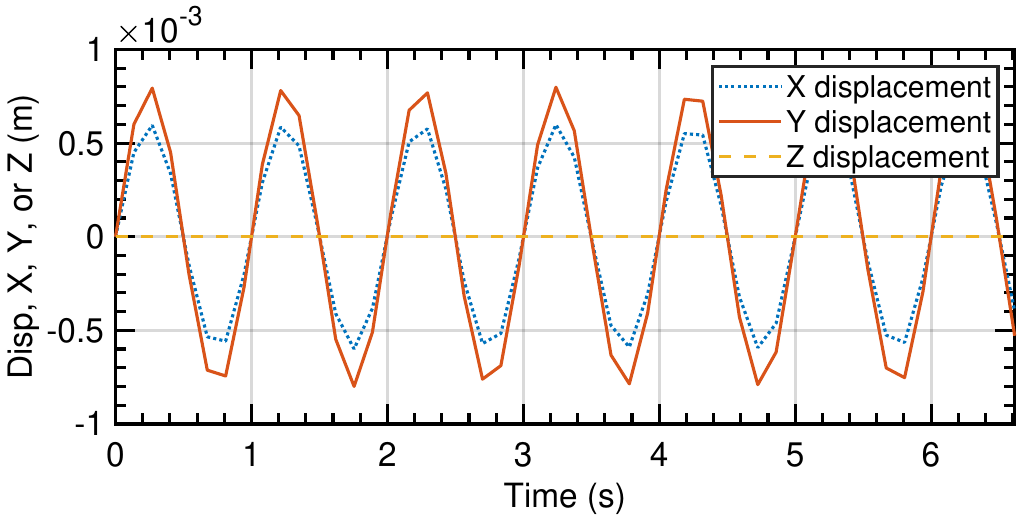}}
\caption{Target displacements (ground truth). Note that the initial position of the target is subtracted for display purposes.}
\label{fig:simulation target movement (ground truth)}
\vspace{0mm}\end{figure}

\begin{figure}
    \centering
    \begin{subfigure}{.45\textwidth}
        \centering
        \includegraphics[width=.95\linewidth]{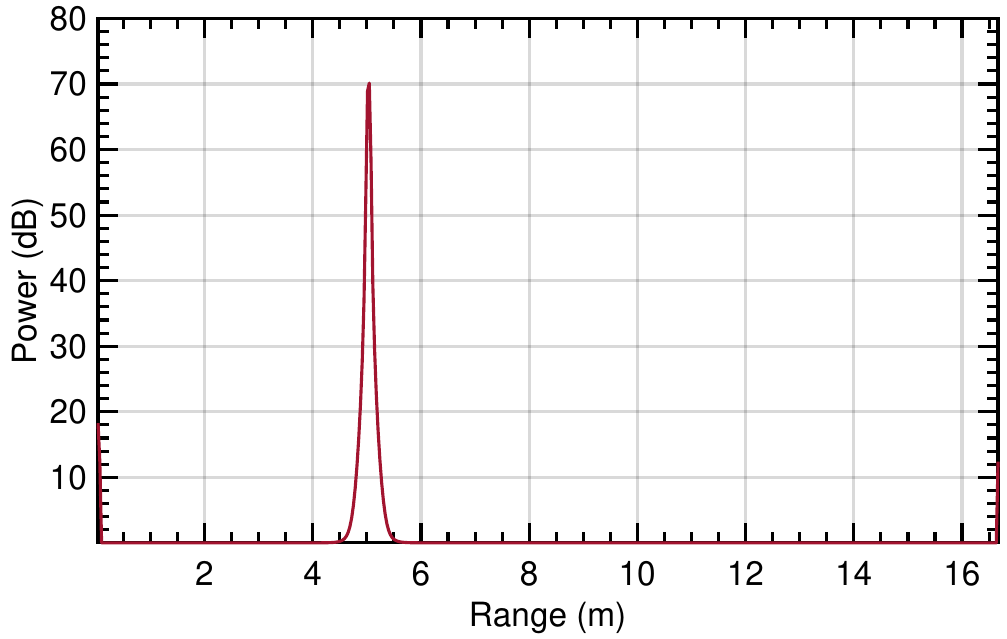} 
        \caption{Range profile}
    \end{subfigure}%
    \hfill
    \begin{subfigure}{.45\textwidth}
        \centering
        \includegraphics[width=.95\linewidth]{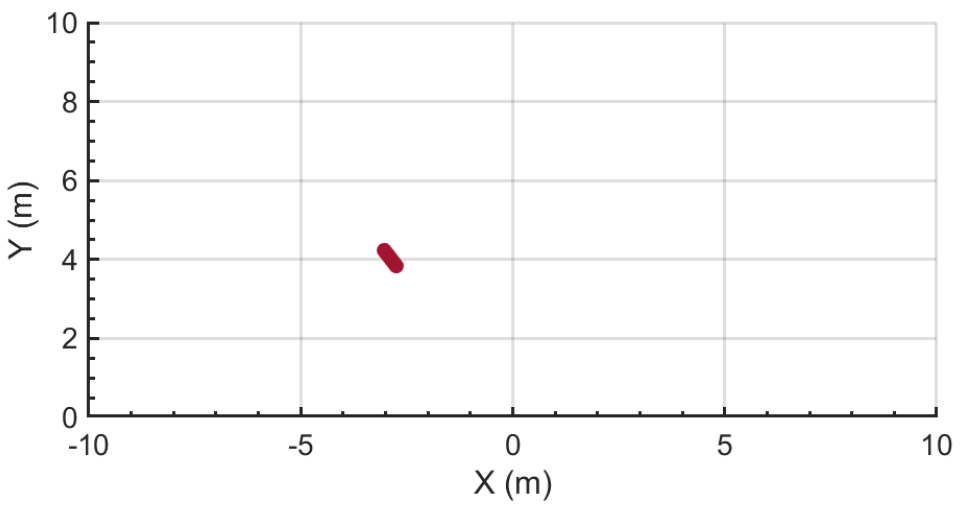} 
        \caption{3D point cloud (xy view)}
    \end{subfigure}%
    \hfill
    \begin{subfigure}{.45\textwidth}
        \centering
        \includegraphics[width=.95\linewidth]{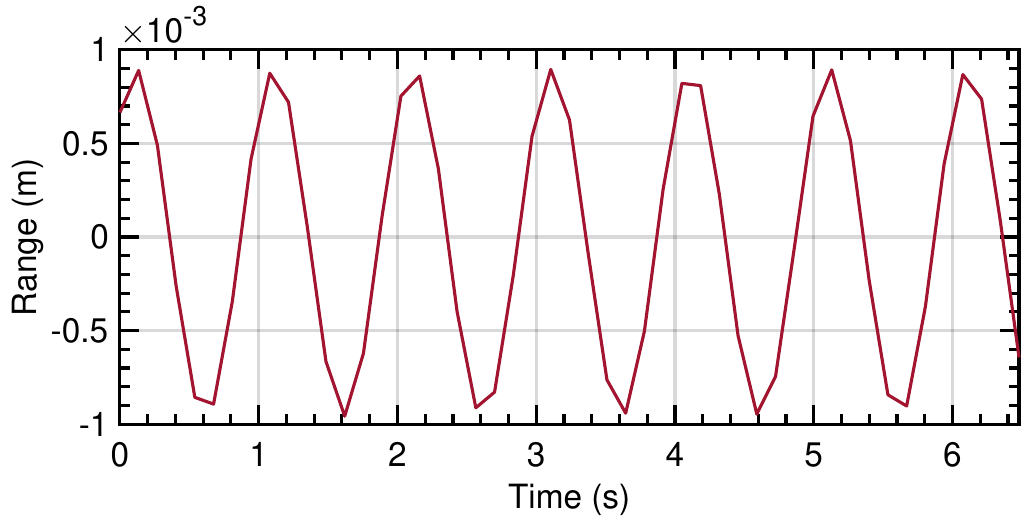} 
        \caption{Estimated target phase signal}
    \end{subfigure}
    \caption{Estimated results for simulation.}
    \label{fig:estimated results for simulation}
\end{figure}

To validate the correctness and validity of the radar signal processing module as described in Section~\ref{sec:radar signal processing} and Section~\ref{subsec:overall processing framework}, we conducted the following simulation. 

In this simulation, the mmWave MIMO Radar parameters are set as Table.~\ref{table:radar parameters}. 
With this parameter setting, the bandwidth is $
B = {{K_{\text{Chirp}}N_{\text{ADCSamples}}}/f_{s}} = 4.61~\text{GHz}$ and range resolution is $\Delta R = {c/{2B}} = 3.25~\text{cm}$. 
For antenna array positions, we use the parameters of TI AWR2243 Cascade Radar RF Board \cite{TI_MMWCAS_RF_EVM}, as shown in Fig.~\ref{fig:antenna array positions}. 
This radar is equipped with 12 transmitting antennas and 16 receiving antennas. 
\revised{Actually, in this research of multi-point vital sign monitoring, the most important parameters that influence the performance are carrier frequency $f_c$ of radio waveform and angular resolutions including azimuth and elevation, i.e., $\Delta \theta_{\text{azi}}$ and $\Delta \theta_{\text{ele}}$. The carrier frequency $f_c$ is determined by the hardware system and can be set as any value between $76~\text{GHz}$ and $81~\text{GHz}$. And there is no great difference for different values of carrier frequencies since their wavelengths are very close. 
Angular resolutions depend on the number of formed virtual antennas. And we used all the available antennas on the radio frequency (RF) board.}
With this antenna array setting, the steering vectors for transmitting antenna array and receiving antenna array in (\ref{eq:transmitting vector a}) and (\ref{eq:receiving vector b}) are expressed as in (15) and (16), respectively. 
As is shown in Fig.~\ref{fig:simulation target movement (ground truth)}, the simulation target is at the position of $
\left\lbrack {3,~4,~0} \right\rbrack~\text{m}$, in a sinusoidal movement with the frequency as $1.0~\text{Hz}$ and the amplitude as $
\left\lbrack {0.6,~0.8,~0} \right\rbrack~\text{mm}$, i.e., $1.0~\text{mm}$ as the composite movement. 

From the estimated results in Fig.~\ref{fig:estimated results for simulation}, we can see that using the radar signal processing module, the target position (range and angle) and displacement can be accurately estimated. 
The estimated target displacement is also sinusoidal, with an amplitude of approximately $1.0~\text{mm}$ ($6.44~\text{rad}$).  
It is important to note that the radar estimates the composite displacement of the target.
These results demonstrate that our signal processing steps are effective to estimate the target displacement.

\vspace{-3mm}\begin{equation}
\mathbf{a}\left( {\theta_{k},\phi_{k}} \right)= \begin{bmatrix}
{e^{ {j2\pi\left( {5.5{\cos\phi_{k}}{\sin\theta_{k}} +3.0{\sin\phi_{k}}{\sin\theta_{k}}} \right)} }} \\
{e^{{j2\pi\left( {5.0{\cos\phi_{k}}{\sin\theta_{k}} + 2.0{\sin\phi_{k}}{\sin\theta_{k}}} \right)} }} \\
{e^{ {j2\pi\left( {4.5{\cos\phi_{k}}{\sin\theta_{k}} + 0.5{\sin\phi_{k}}{\sin\theta_{k}}} \right)} }} \\
{e^{ {j2\pi\left( {16{\cos\phi_{k}}{\sin\theta_{k}}} \right)} }} \\
{e^{{j2\pi\left( {14{\cos\phi_{k}}{\sin\theta_{k}}} \right)} }}\vspace{-1.5mm} \\
 \vdots \\
{e^{{j2\pi\left( {2{\cos\phi_{k}}{\sin\theta_{k}}} \right)} }} \\
1
\end{bmatrix}_{12 \times 1}
\end{equation}\vspace{-3mm}

\begin{equation}
\vspace{-6mm}
\mathbf{b}\left( {\theta_{k},\phi_{k}} \right) = \begin{bmatrix}
{e^{ {- j2\pi\left( {\left( {5.0 + 0.5 \times 1} \right){\cos\phi_{k}}{\sin\theta_{k}}} \right)} }} \vspace{-2mm}\\
 \vdots \\
{e^{ {- j2\pi\left( {\left( {5.0 + 0.5 \times 4} \right){\cos\phi_{k}}{\sin\theta_{k}}} \right)} }} \\
{e^{ {- j2\pi\left( {\left( {22.5 + 0.5 \times 5} \right){\cos\phi_{k}}{\sin\theta_{k}}} \right)} }} \vspace{-2mm}\\
 \vdots \\
{e^{ {- j2\pi\left( {\left( {22.5 + 0.5 \times 8} \right){\cos\phi_{k}}{\sin\theta_{k}}} \right)} }} \\
{e^{ {- j2\pi\left( {\left( {18.5 + 0.5 \times 9} \right){\cos\phi_{k}}{\sin\theta_{k}}} \right)} }} \vspace{-2mm}\\
 \vdots \\
{e^{ {- j2\pi\left( {\left( {18.5 + 0.5 \times 12} \right){\cos\phi_{k}}{\sin\theta_{k}}} \right)} }} \\
{e^{  {- j2\pi\left( {\left( {- 6.5 + 0.5 \times 13} \right){\cos\phi_{k}}{\sin\theta_{k}}} \right)} }}\vspace{-2mm} \\
 \vdots \\
{e^{  {- j2\pi\left( {\left( {- 6.5 + 0.5 \times 16} \right){\cos\phi_{k}}{\sin\theta_{k}}} \right)} }}
\end{bmatrix}_{16 \times 1}
\end{equation}

\vspace{2mm}\section{Experiment and Discussion}
\label{sec:experiment}

In this section, we first introduce the experimental setup, including the sensor configurations and the spatial alignment scheme for the radar system and the SCG sensors. 
After that, we present the outcomes of the dual reference modalities. 
Following this, we present the results obtained by the radar system in different postures and breathing patterns, comparing them to those from the reference modalities.
Lastly, we discuss the limitations of the proposed MultiVital system and possible solutions.  

\subsection{Experimental Setting}
\begin{figure}[htbp]
\centerline{\includegraphics[scale=0.33]{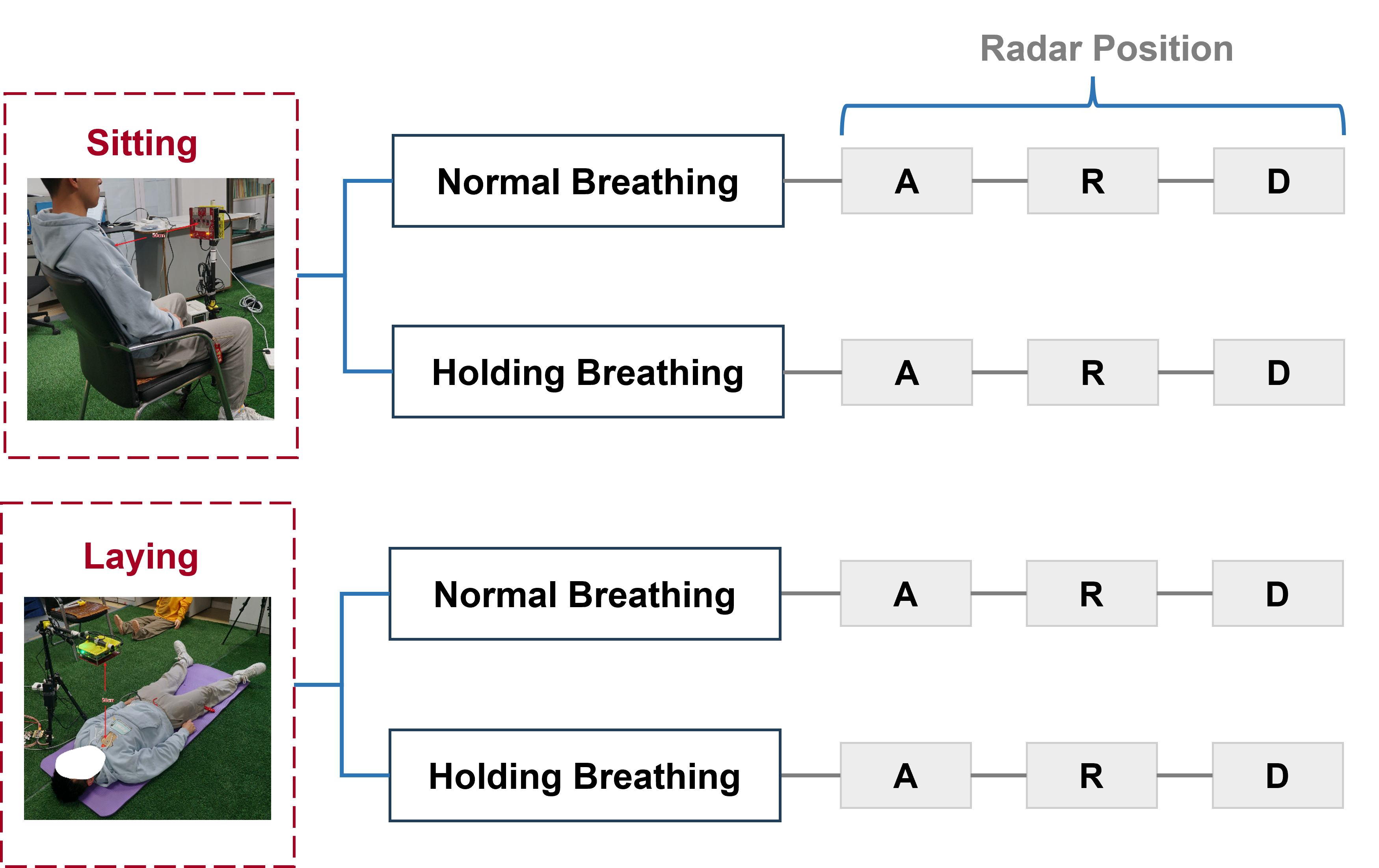}}
\caption{\revised{Twelve experimental groups with two different postures (sitting and laying), two different breathing patterns (normal and holding), and three radar positions (A: radar directed at point A, R: radar positioned to the right, and D: radar positioned below).}}
\label{fig:experiment_scenario}
\end{figure}

\revised{To comprehensively capture cardiac signals from the human body across various postures, breathing patterns, and radar positions, we designed the following experimental groups
\footnote{The informed consent was obtained from all subjects in the experiments.}}. 
\revised{
As shown in Fig.~\ref{fig:experiment_scenario}, the experiments were first categorized based on the posture of the human body, specifically into sitting and laying groups. 
Each posture group was further subdivided according to breathing patterns—normal breathing and holding breathing—along with three radar positions for each posture group, resulting in a total of 12 experimental groups.}
\revised{
In each experimental group, three subjects participated, each undergoing three trials per group. 
The quantitative data presented in this section were calculated by averaging the results across the nine total trials for each experimental group.
}

\revised{
It is worth noting that in most clinical settings, the relative position between the radar and the human body is generally fixed, although minor relative movements may occur. 
To experimentally investigate the influence of radar position on detection performance, we tested three configurations in our experiments: (A) the radar antenna array centered directly to point A, (R) offset by 5 cm to the right, and (D) offset by 5 cm downward.
}

\begin{figure}[t]
\vspace{-8mm}
    \centering
    \begin{subfigure}{0.5\columnwidth}
        \centering
        \includegraphics[width=.95\linewidth]{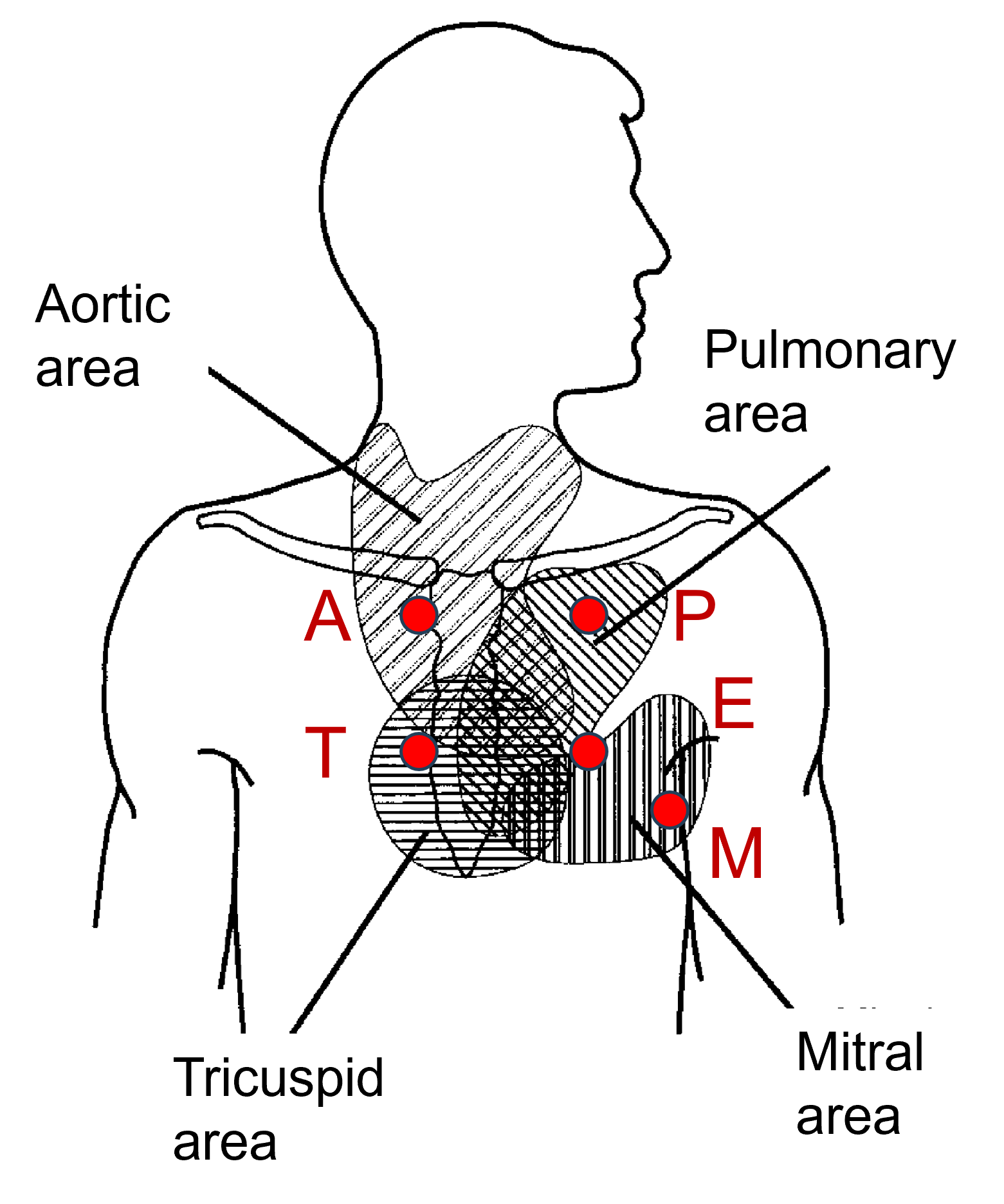} 
        \caption{\revised{SCG sensors}}
    \end{subfigure}%
    \begin{subfigure}{0.5\columnwidth}
        \centering
        \includegraphics[width=.85\linewidth]{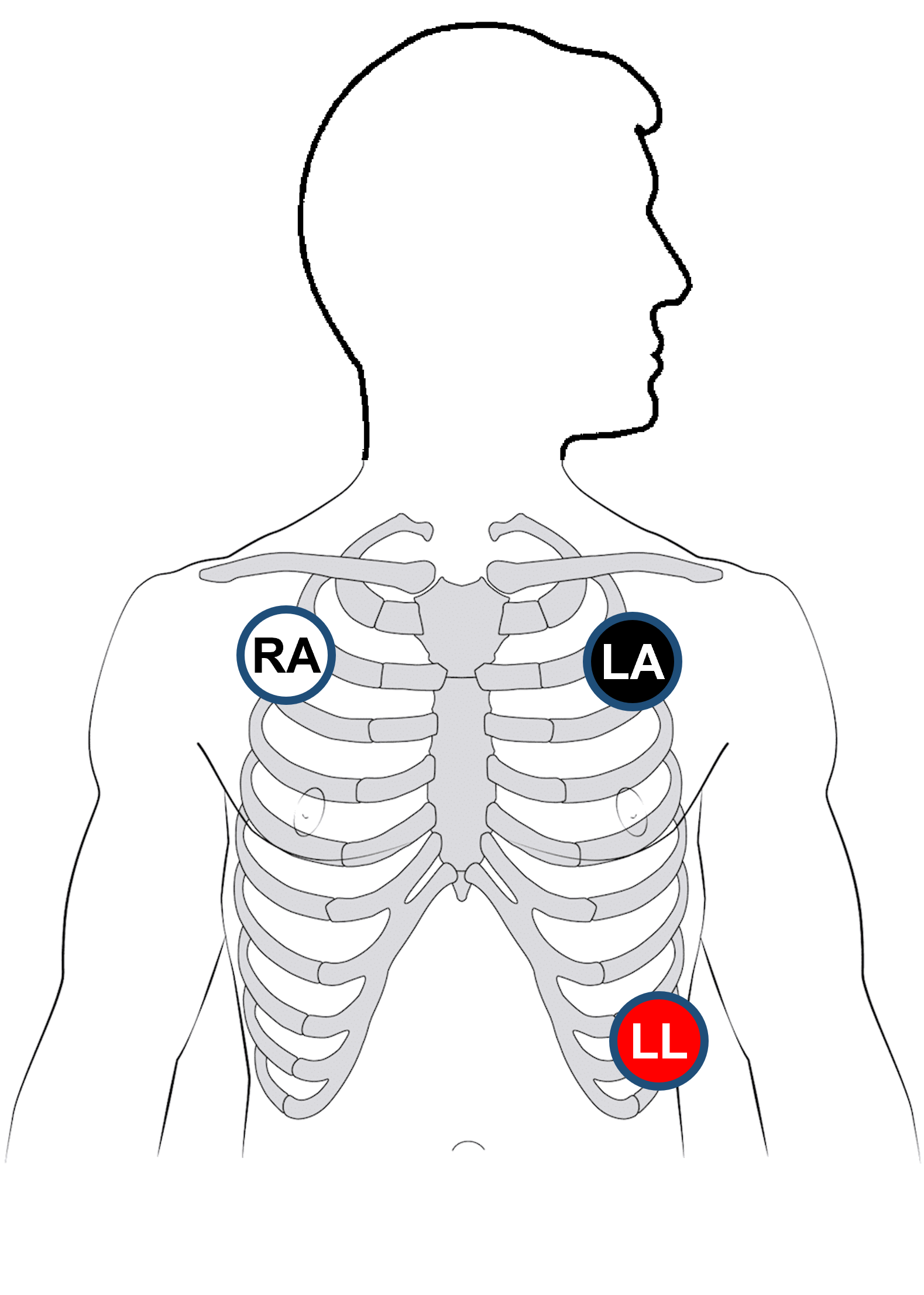} 
        \caption{\revised{ECG eletrodes}}
    \end{subfigure}%
    \hfill
    \begin{subfigure}{0.5\columnwidth}
        \centering
        \includegraphics[width=.95\linewidth]{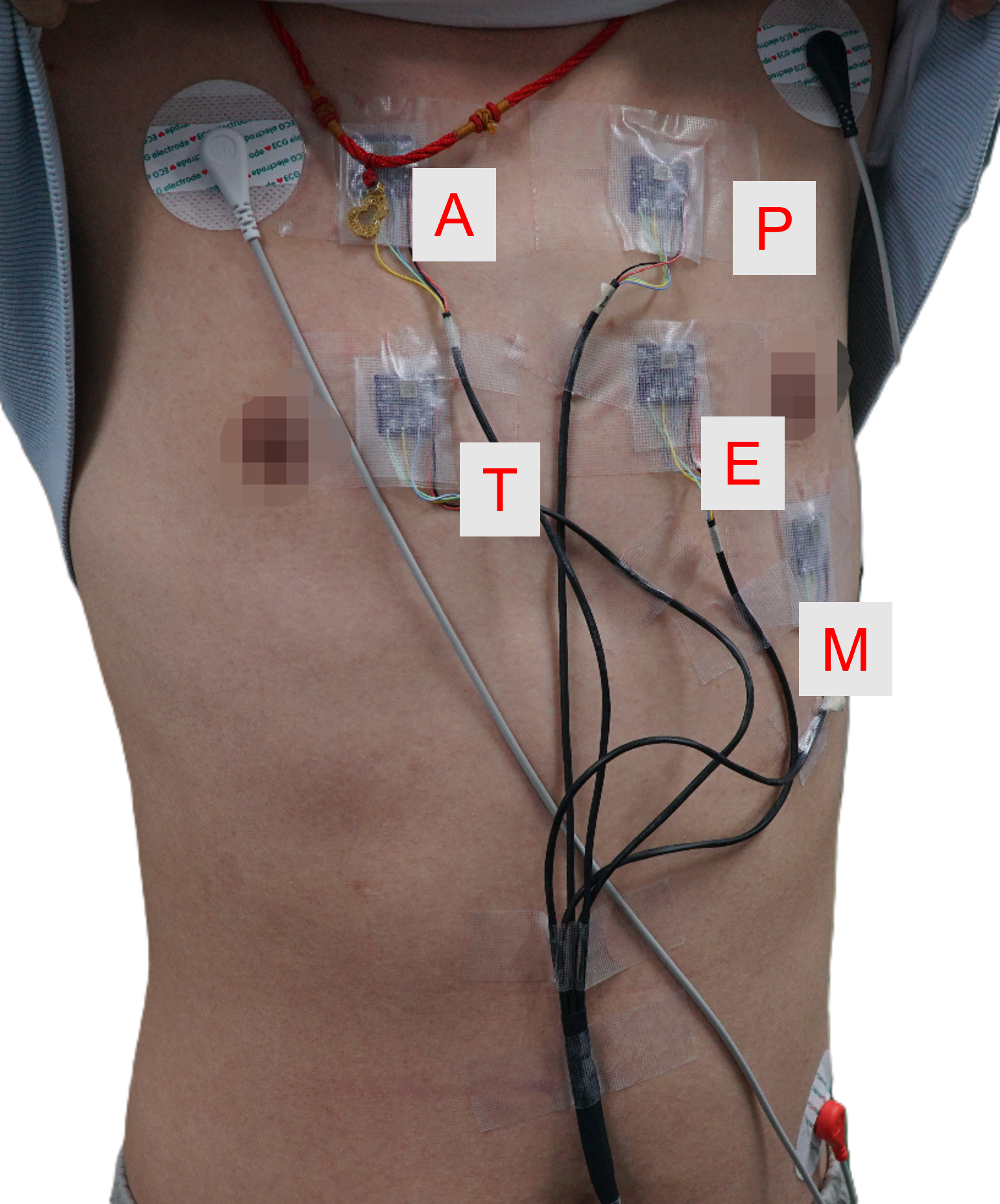} 
        \caption{Real deployment}
    \end{subfigure}%
    \caption{Positions of SCG sensors and ECG eletrodes. (a) The placement positions of five SCG sensors on the surface of the chest, where regions 1-5 correspond to the A, P, T, E, M areas of the heart auscultation zones, respectively. (b) The placement positions of the three-lead ECG electrodes on the surface of the torso, where the RA electrode is located at the right shoulder of the subject, the LA electrode is placed at the left shoulder of the subject, and the LL electrode is placed on the left side of the subject's waist. (c) Real positions of SCG sensors and ECG eletrodes.}
    \label{fig:positions of SCG sensors and ECG eletrodes}
\end{figure}

The duration of each group’s experiment was set as 3 minutes. 
Normal breathing refers to the subject breathing steadily according to their natural rhythm under normal conditions. 
Holding breathing involves the subject holding their breath for the first 30 seconds before data collection, maintaining the chest without significant undulation, followed by normal breathing in the next 30 seconds, and so on, in a cycle until 3 minutes are completed. 
\revised{In the experiment, we aim to measure five points that are realted to cardiac auscultation, i.e., A, P, T, E, M, as shown in Fig.~\ref{fig:positions of SCG sensors and ECG eletrodes}. In medical field, the aortic (A), pulmonary (P), mitral (M), and tricuspid (T) areas are of interest as different pathologies can be felt at each location \cite{taylor2015learning}. Also considering the spatial symmetry of the multi-point arrangement and the convenience of the experiment, we chose these five measurement points of interest.}

mmWave MIMO radar setting. 
To ensure the mmWave MIMO radar could capture accurate phase change data and separate multiple chest regions, 12 Tx and 16 Rx antennas were activated and used, forming a 2D antenna array composed of 192 channels.
This array employs a Time Division Multiplexing (TDM) mode for millimeter-wave transmission. The specific setting parameters are shown in Table.~\ref{table:radar parameters for experiment}. 
With this parameter setting, the total effective bandwidth is $3.38~\text{GHz}$ and range resolution is $4.44~\text{cm}$. 
\revised{In addition, the number of chirps in one frame, $N_{\text{ChirpsOneFrame}}$, is set as one to reduce the data amount.}
\revised{Moreover, in experimental scenarios, the far field condition is not satisfied, and near-field calibration is incorporated, introduced in detail in Appendix.}

\begin{table}[htbp]
\centering
\caption{mmWave MIMO radar parameters for experiment.}
\label{table:radar parameters for experiment}
\begin{tabular}{l|l|l}
\toprule
\textbf{Parameter Name}       & \textbf{Symbol}   & \textbf{Value} \\
\midrule
Carrier Frequency             & $f_c$              & $77~\mathrm{GHz}$         \\
Pulse Repetition Time         & PRT               & $70~\mu\mathrm{s}$        \\
Frame Duration                & $T_{\text{Frame}}$          & $50~\mathrm{ms}$        \\
Number of ADC Samples         & $N_{\text{ADCSamples}}$     & $256$            \\
Sampling   Frequency          & $f_s$              & 5 MHz          \\
Chirp Rate                    & $K_{\text{Chirp}}$          & $65.998~\mathrm{MHz}/\mu\mathrm{s}$  \\
Number of Chirps in One Frame & $N_{\text{ChirpsOneFrame}}$ & $1$            \\
Total Frames                  & $N_{\text{Frame}}$          & $3600$             \\
\bottomrule
\end{tabular}
\end{table}

\vspace{-3mm}\begin{table}[htbp]
\centering
\caption{Setting for SCG sensor channels.}
\label{table:setting for acceleration sensors}
\begin{tabular}{m{2.5cm}|m{2cm}}
\toprule
\textbf{Parameter Name}       & \textbf{Value} \\
\midrule
Sampling Frequency              & $20~\mathrm{ksps}$         \\
Measurement Range               & $-2~\text{V to }+2~\text{V}$        \\
Sampling Depth                  & $16~\mathrm{Bit}$        \\
\bottomrule
\end{tabular}
\end{table}

Multiple SCG (acceleration sensor) and ECG setting. 
Five SCG sensors were placed at the positions shown in Fig.~\ref{fig:positions of SCG sensors and ECG eletrodes}(a) and \ref{fig:positions of SCG sensors and ECG eletrodes}(c), and the ECG used the typical placement for a three-lead ECG, with specific placement shown in Fig.~\ref{fig:positions of SCG sensors and ECG eletrodes}(b) and \ref{fig:positions of SCG sensors and ECG eletrodes}(c). 
Data collection and recording were conducted using LabChart software, with 16 channels enabled and configured on the software. 
This included 15 acceleration sensor channels and one ECG channel. 
Specific parameters were set according to Table.~\ref{table:setting for acceleration sensors}, where the ECG measurement range was from $-5~\text{mV to }+5~{\text{mV}}$. 
Other parameter settings were consistent with those of the acceleration sensor channels.

Spatial alignment scheme of mmWave MIMO radar and five-group SCG sensors. 
To use SCG measured results as the reference results for verifying the accuracy of mmWave MIMO radar outcomes, it is necessary to precisely locate the five SCG measurement points—A, P, T, E, M—within the radar processing results. To this end, we adopt the following alignment scheme:
\begin{itemize}
    \item Place SCG sensors at the five point A, P, T, E, M as shown in Fig.~\ref{fig:spatial alignment scheme} and measure the positions of A, P, T, E, M, denoted as $\left\lbrack {x_{A},y_{A}} \right\rbrack$, $\left\lbrack {x_{P},y_{P}} \right\rbrack$, $\left\lbrack {x_{T},y_{T}} \right\rbrack$, $\left\lbrack {x_{E},y_{E}} \right\rbrack$, and $\left\lbrack {x_{M},y_{M}} \right\rbrack$, respectively.
    \item Align the geometric center of the radar antenna array with point A, that is, positioning point A at the location where the radar’s azimuth and elevation angles are zeros, denoted as $\left\lbrack {\phi_{A}^{\star} = 0,\theta_{A}^{\star} = 0} \right\rbrack$.
    \item From the radar processed results, estimate the radial distance $z_A$ from point A to the radar and calculate the ground-truth azimuth and elevation angles for points P, T, E, and M relative to the radar, denoted as $\left\lbrack {\phi_{P}^{\star},\theta_{P}^{\star}} \right\rbrack$, $\left\lbrack {\phi_{T}^{\star},\theta_{T}^{\star}} \right\rbrack$, $\left\lbrack {\phi_{E}^{\star},\theta_{E}^{\star}} \right\rbrack$, $\left\lbrack {\phi_{M}^{\star},\theta_{M}^{\star}} \right\rbrack$, where $\phi_{i}^{\star} = {\arctan\left( {\left( x_{i} - x_{A} \right)/\left| z_{A} \right|} \right)}$ and $\theta_{i}^{\star} = {\arctan\left( {\left( \theta_{i} - \theta_{A} \right)/\left| z_{A} \right|} \right)}$ for $i=P,T,E,M$.
    \item Based on the ground-truth azimuth and elevation angles of each point, identify the corresponding signals for each point from the radar processed results.
\end{itemize}


\begin{figure*}[htb]
\centering
\subcaptionbox{\revised{$x$-axis, $y$-axis, and $z$-axis accelerations of point A.}}{\includegraphics[width=0.48\linewidth]{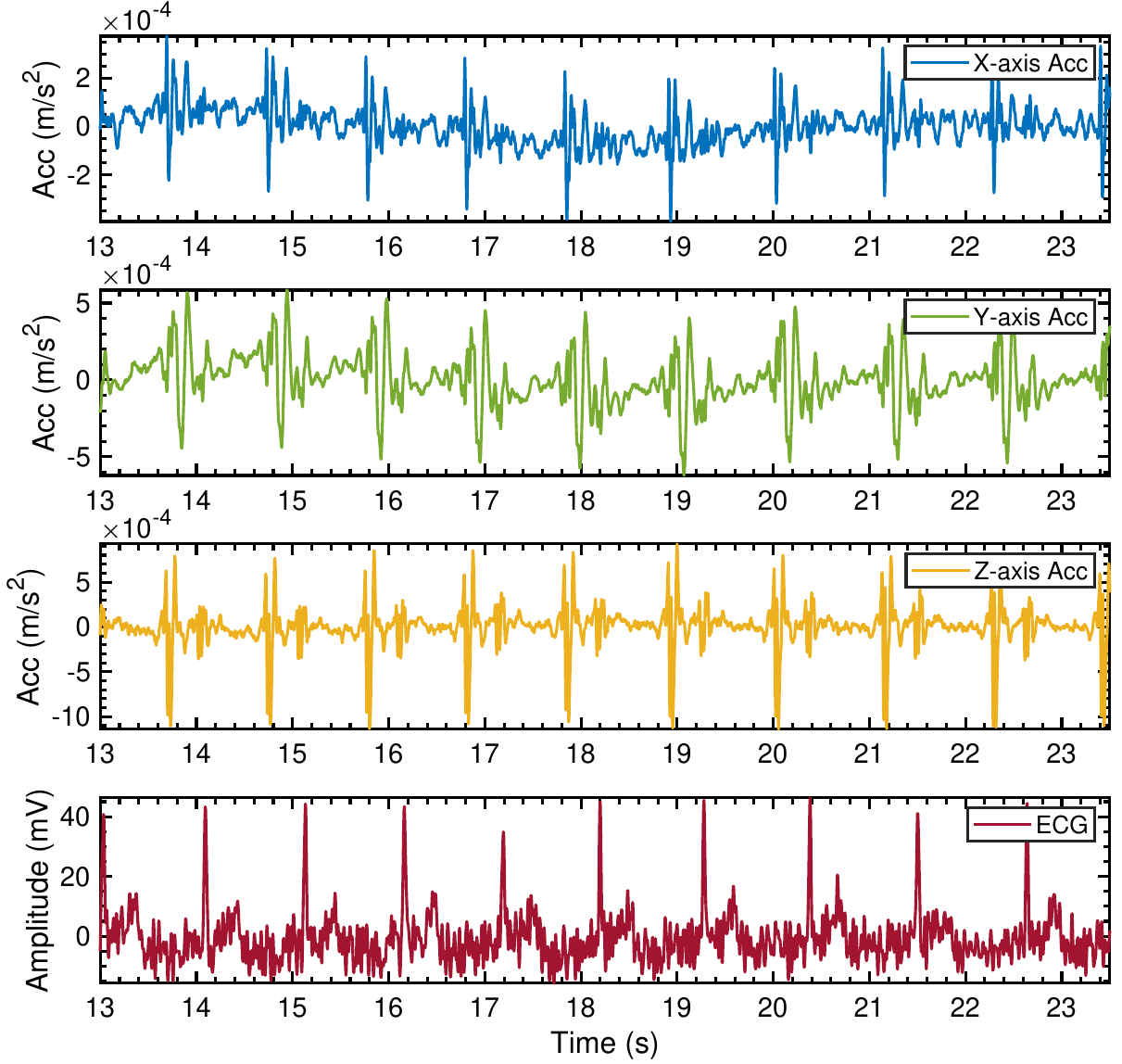}}
\hfill
\subcaptionbox{\revised{$x$-axis, $y$-axis, and $z$-axis displacements of point A.}}{\includegraphics[width=0.48\linewidth]{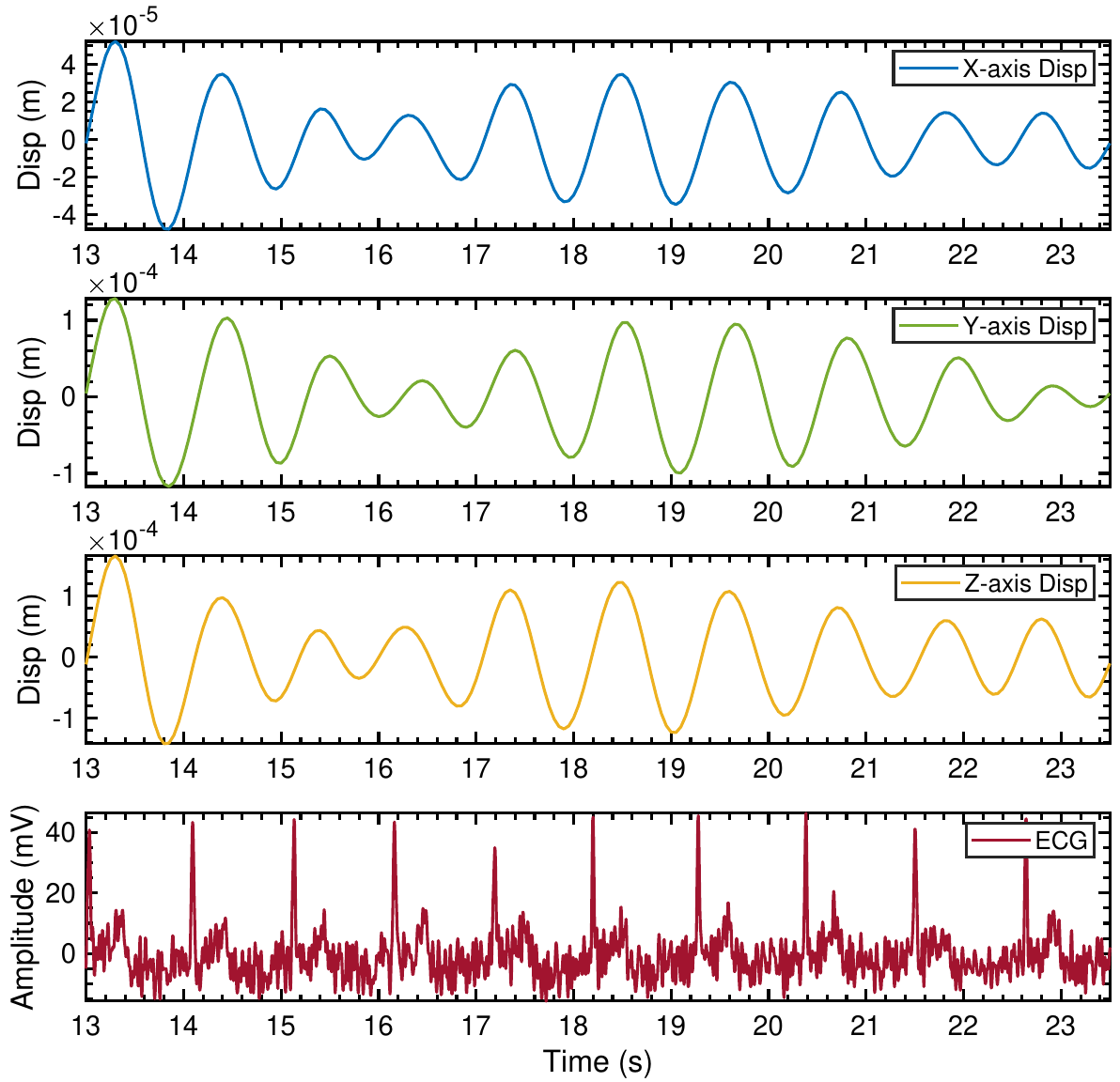}}
\caption{\revised{$x$-axis, $y$-axis, and $z$-axis accelerations of point A measured by SCG sensors and the calculated displacements. The ECG is also presented for comparison. The human subject is in laying posture and holding breathing pattern.}}
\label{fig:xyz accelerations, velcities, and displacements of point A measured by SCG sensors} 
\end{figure*}


\begin{figure*}[htb]
\centering
\begin{minipage}[t]{0.48\linewidth}
\centering
\includegraphics[width=\linewidth]{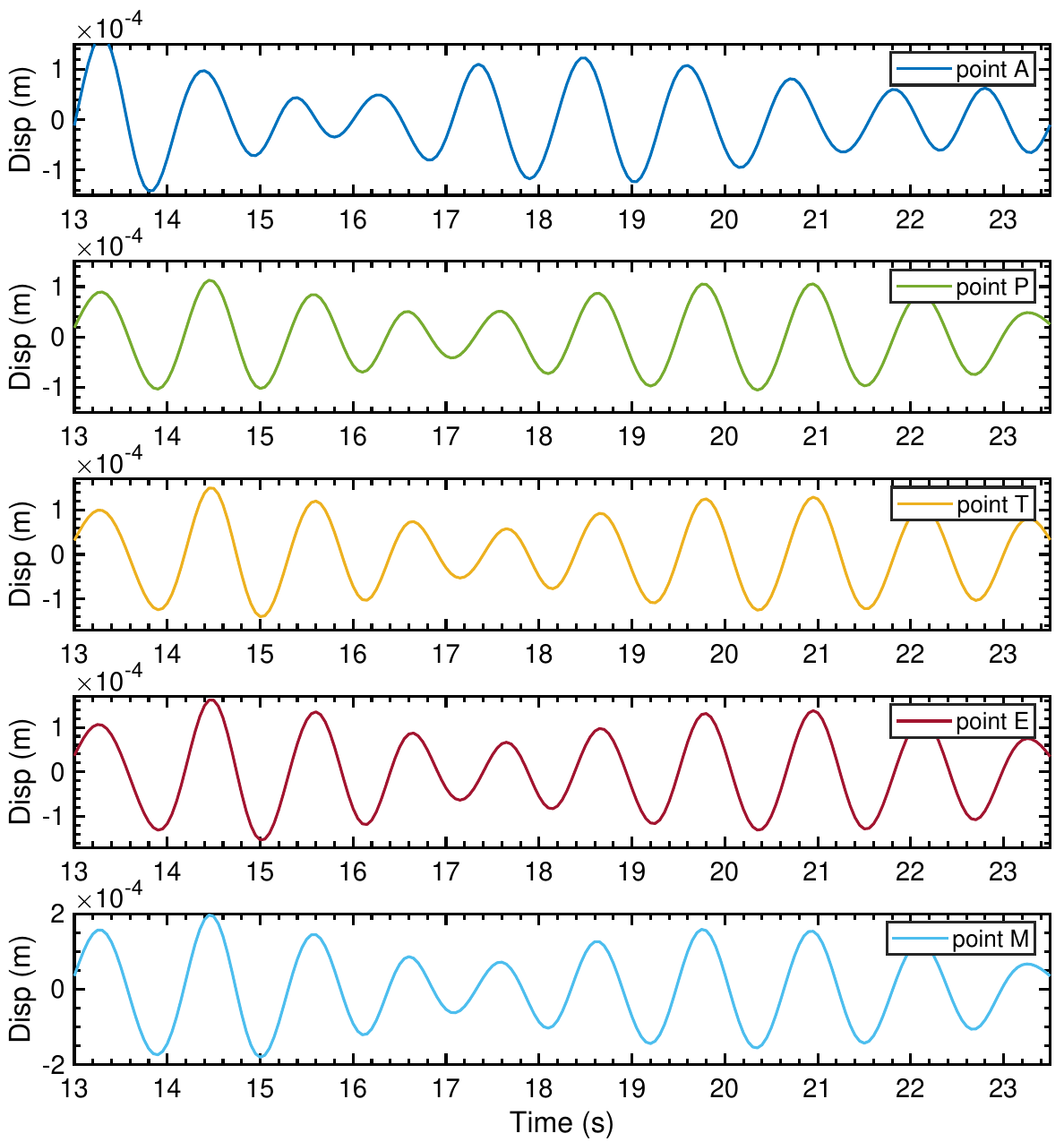}
\caption{\revised{$z$-axis displacements calculated using the corresponding accelerations measured by SCG sensors. The human subject is in laying posture and holding breathing pattern.}}
\label{fig:xyz displacements of the five points calculated using SCG data}
\end{minipage}\hfill
\begin{minipage}[t]{0.48\linewidth}
\centering
\includegraphics[width=\linewidth]{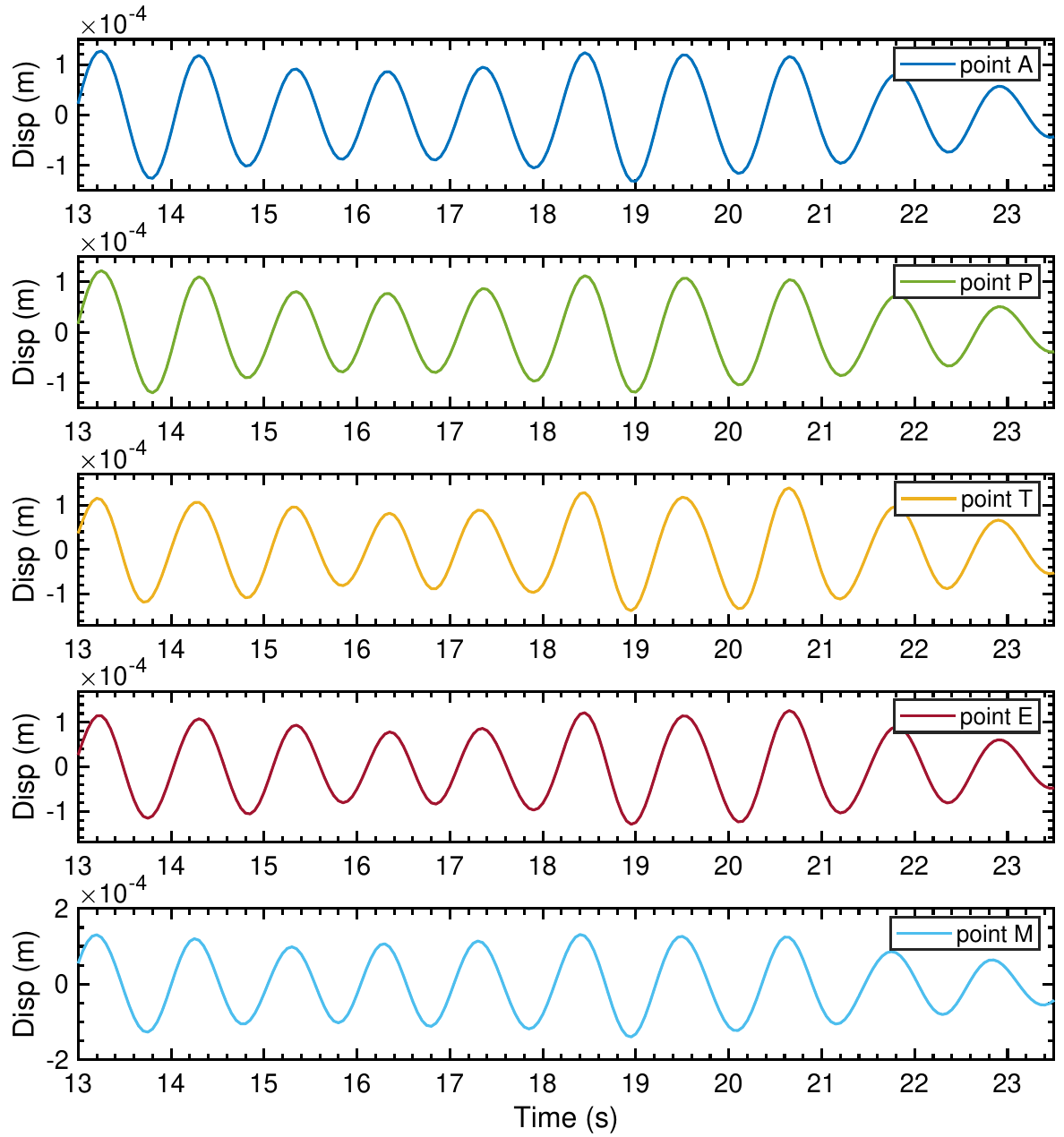}
\caption{\revised{Radar estimated five-point displacements for laying posture and holding breathing pattern.}}
\label{fig:radar estimated five-point displacements}
\end{minipage}
\end{figure*}


\subsection{Results of Dual Reference Modalities}
\label{subsec:SCG results}

As shown in Fig.~\ref{fig:overall processing framework}, the ECG data can be directly used as one reference result without any further processing. The last sub-figures in Fig.~\ref{fig:xyz accelerations, velcities, and displacements of point A measured by SCG sensors} present the ECG results of point A for laying posture and holding breathing pattern.

The 5 SCG sensors (15 accelerators) monitor the five regions of interest at the same time as the mmwave MIMO radar. 
And then the results obtained using SCG sensors are provided as the reference results to test the performance of radar estimated results. 
Each SCG sensor can estimate the displacement of the interested region by measuring the $x$-axis, $y$-axis, and $z$-axis accelerations and calculating the displacements.

Fig.~\ref{fig:xyz accelerations, velcities, and displacements of point A measured by SCG sensors} presents the SCG results of point A for laying posture and holding breathing pattern, including the measured $x$-axis, $y$-axis, and $z$-axis accelerations and the calculated displacements. 
As shown in Fig.~\ref{fig:xyz accelerations, velcities, and displacements of point A measured by SCG sensors}(a), the measured acceleration results are highly consistent with the ECG data in periodicity. 
The peak-to-peak differences of the calculated $x$-axis, $y$-axis, and $z$-axis displacements are approximately from \revised{ $0.1~\text{mm}$ to $0.3~\text{mm}$ }, which is shown in Fig.~\ref{fig:xyz accelerations, velcities, and displacements of point A measured by SCG sensors}(b). 

Fig.~\ref{fig:xyz displacements of the five points calculated using SCG data} presents the \revised{$z$}-axis displacements of the five points of interest calculated using the corresponding accelerations measured by SCG sensors. 
The human subject is in laying posture and holding breathing pattern. 
Fig.~\ref{fig:xyz displacements of the five points calculated using SCG data} shows that the displacements of the five interested points are highly consistent with each other in periodicity, with slight differences. 

\begin{figure}[htbp]
\centerline{\includegraphics[scale=0.4]{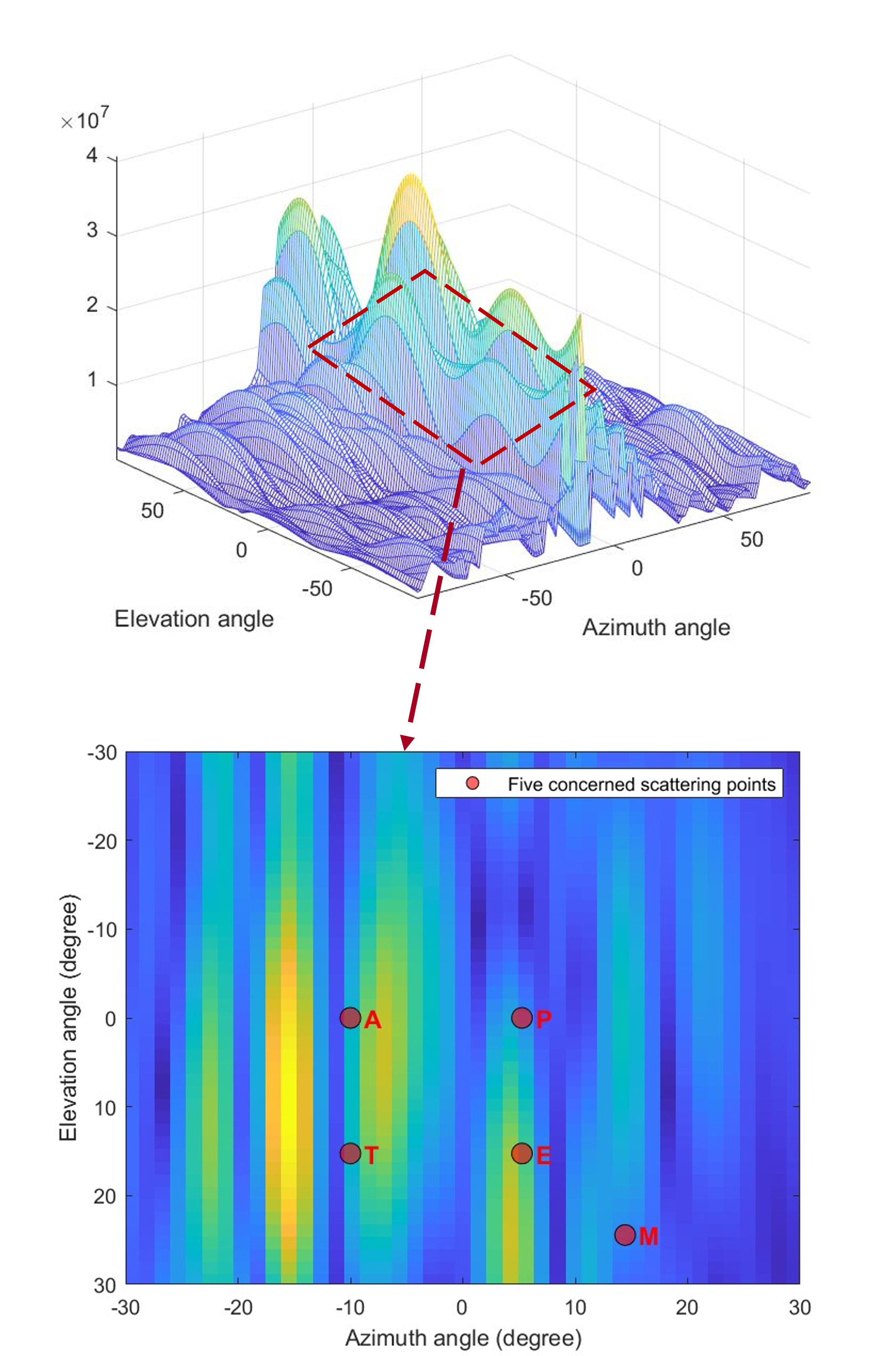}}
\caption{DOA processing results in the range bin where the chest wall of the human subject is located.}
\label{fig:DOA processing results}
\end{figure}


\begin{figure}[htbp]
\vspace{-3mm}\centerline{\includegraphics[scale=0.48]{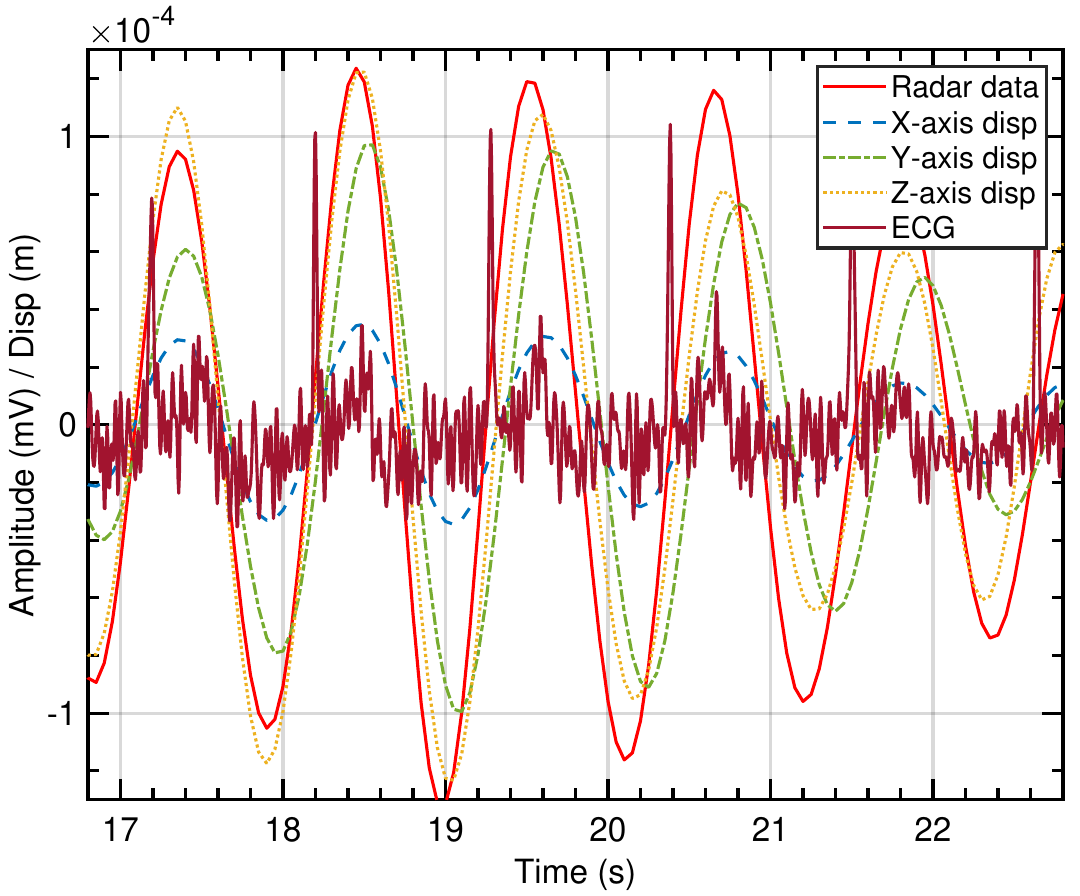}}
\caption{\revised{Comparison of radar result, $x$-axis, $y$-axis, and $z$-axis displacements of point A estimated with SCG data, and ECG signal. Note that ECG signal is scaled appropriately to fit plotting needs.}}
\label{fig:comparison of radar, SCG, and ECG signal}
\end{figure}

\begin{table}[htbp]
\centering
\caption{Comparison of radar result and $x$-axis, $y$-axis, and $z$-axis displacements of point A measured by SCG sensors in the experiment scenario of laying posture and holding breathing pattern.}
\label{table:comparison of radar and SCG at point A}
\begin{tabular}{m{1.0cm}|m{3cm}|m{3cm}}
\toprule
\textbf{}      & \textbf{Maximum value of cross-correlation},$\rho_{rx}$ & \textbf{Difference in maximum frequency},$(\Delta f_{rx})_{\text{max}}$,~Hz \\
\midrule
Radar \& SCG-X & \revised{0.88}                                        & 0                                        \\
Radar \& SCG-Y & \revised{0.84}                                        & 0                                        \\
Radar \& SCG-Z & \revised{0.87}                                        & 0                                        \\
\bottomrule
\end{tabular}
\end{table}

\subsection{Results of mmWave MIMO Radar}
In this subsection, we take the experimental results of laying posture and holding breathing pattern as an example to analyze the performance of our MultiVital solution.

For the data of each frame, after DOA processing, we can fuse the signals of multiple channels to extract target spatial information, that is, the target information at different azimuth and different elevation angles. 
Then we can locate the positions of the five points with the aforementioned ground-truth azimuth and elevation angles, i.e., $\left\{ \left\lbrack {\phi_{h}^{\star},\theta_{h}^{\star}} \right\rbrack \right\}_{h = A,P,T,E,M}$. 
Fig.~\ref{fig:DOA processing results} shows the DOA processed result and the five points of interest. 
Then the slow-time phase signals of each point can be obtained by stacking the data from all frames.

\revised{Fig.~\ref{fig:radar estimated five-point displacements} presents the estimated displacements of the five points of interest using radar, which are derived from the slow-time phase signals of these points.} 
The following observations can be drawn from the results:
\begin{itemize}
    \item The five-point (A, P, T, E, and M) signals are generally identical with slight differences.  This observation is consistent with the fact that the five different regions on the human chest wall present very similar displacements with slight difference.
    \item \revised{The estimated displacements with radar closely match those obtained from the reference multiple SCG sensors, demonstrating the accuracy of radar solutions.}
\end{itemize}

In order to quantitatively compare the two signals $r(t)$ and $x(t)$, we take two metrics: 
1) maximum value of cross-correlation, $\rho_{rx}$. $\rho_{rx}$ can be calculated using (\ref{eq:rho_rx}) and (\ref{eq:C_rx}), 
2) the difference in maximum frequency, $(\Delta f_{rx})_{\text{max}}$.

\begin{equation} \label{eq:rho_rx}
\rho_{rx} = {\max\limits_{m}\left| {C_{rx}(m)} \right|},
\end{equation}
\begin{equation} \label{eq:C_rx}
C_{rx}(m) = \left\{ \begin{matrix}
{{\sum_{n = 0}^{N - m - 1}{\overset{\sim}{r}\left( {n + m} \right){\overset{\sim}{x}}_{n}^{*}}},} & {m \geq 0} \\
{C_{xr}^{*}\left( {- m} \right)}, & {m < 0}
\end{matrix} \right.,
\end{equation}
where $\overset{\sim}{r}(n)$ and  $\overset{\sim}{x}(n)$ are normalized $r(n)$ and $x(n)$, respectively.
\begin{equation}
\left( {\Delta f_{rx}} \right)_{\text{max}} = \left| {\left( f_{r} \right)_{\text{max}} - \left( f_{x} \right)_{\text{max}}} \right|,
\end{equation} 
where  $(f_r)_{\text{max}}$ and $(f_x)_{\text{max}}$ denote the maximum frequencies of $r(n)$ and $x(n)$, respectively. 
$|\cdot|$ denotes the absolute operation.


Fig.~\ref{fig:comparison of radar, SCG, and ECG signal} presents the radar result and $x$-axis, $y$-axis, and $z$-axis isplacements. 
We can see that the radar estimated displacements are highly synchronized with the SCG measured displacements and ECG signals. 
They display slight differences in signal amplitude. 
Such differences in amplitude between radar result and $x$-axis, $y$-axis, and $z$-axis displacements come from the \revised{mismatch} between the directions of $x$-axis, $y$-axis, and $z$-axis accelerometers and the radial direction of the radar. 
The results estimated from the data measured by any of the $x$-axis, $y$-axis, or $z$-axis accelerometers represent the displacement in a certain direction. 
The radar measures the radial motion of the scatter point in line with the radar. 
This radial direction does not have a fixed relationship with the directions of the $x$-axis, $y$-axis, and $z$-axis accelerations.
Table.~\ref{table:comparison of radar and SCG at point A} compares the radar result and $x$-axis, $y$-axis, and $z$-axis displacements in terms of maximum value of cross-correlation ($\rho_{rx}$) and difference in maximum frequency ($(\Delta f_{rx})_{\text{max}}$). 
We can see that the radar results and $x$-axis, $y$-axis, and $z$-axis displacements have a very high degree of similarity, with the $\rho_{rx}$ being \revised{0.84 to 0.88} and $(\Delta f_{rx})_{\text{max}}$ being 0. 

\subsection{\revised{Influence of Postures, Breathing Patterns, and Radar Positions}}

\begin{table*}[]
\centering
\caption{\revised{Experimental results: maximum values of cross-correlations $\rho_{rx}$ and differences in maximum frequencies $(\Delta f_{rx})_{\text{max}}$ for different postures, breathing patterns, and radar positions.}}
\label{table: experimental results}
\revised{
\begin{tabular}{lllcccccccccc}
\hline
\multirow{2}{*}{\textbf{Posture}} & \multirow{2}{*}{\textbf{\begin{tabular}[c]{@{}l@{}}Breathing \\ Mode\end{tabular}}} & \multirow{2}{*}{\textbf{\begin{tabular}[c]{@{}l@{}}Radar \\ Position\end{tabular}}} & \multicolumn{2}{c}{\textbf{A}}                                                                         & \multicolumn{2}{c}{\textbf{P}}                                                                         & \multicolumn{2}{c}{\textbf{T}}                                                                         & \multicolumn{2}{c}{\textbf{E}}                                                                         & \multicolumn{2}{c}{\textbf{M}}                                                                         \\ \cline{4-13} 
                                  &                                                                                     &                                                                                     & \multicolumn{1}{l}{\textbf{$\rho_{rx}$}} & \multicolumn{1}{l}{\textbf{$(\Delta f_{rx})_{\text{max}}$}} & \multicolumn{1}{l}{\textbf{$\rho_{rx}$}} & \multicolumn{1}{l}{\textbf{$(\Delta f_{rx})_{\text{max}}$}} & \multicolumn{1}{l}{\textbf{$\rho_{rx}$}} & \multicolumn{1}{l}{\textbf{$(\Delta f_{rx})_{\text{max}}$}} & \multicolumn{1}{l}{\textbf{$\rho_{rx}$}} & \multicolumn{1}{l}{\textbf{$(\Delta f_{rx})_{\text{max}}$}} & \multicolumn{1}{l}{\textbf{$\rho_{rx}$}} & \multicolumn{1}{l}{\textbf{$(\Delta f_{rx})_{\text{max}}$}} \\ \hline
\multirow{6}{*}{Sitting}          & \multirow{3}{*}{Normal}                                                             & A                                                                                   & 0.64                                     & 0.03                                                        & 0.66                                     & 0.00                                                        & 0.72                                     & 0.00                                                        & 0.72                                     & 0.00                                                        & 0.58                                     & 0.04                                                        \\
                                  &                                                                                     & R                                                                                   & 0.87                                     & 0.00                                                        & 0.66                                     & 0.00                                                        & 0.73                                     & 0.00                                                        & 0.74                                     & 0.00                                                        & 0.66                                     & 0.03                                                        \\
                                  &                                                                                     & D                                                                                   & 0.72                                     & 0.00                                                        & 0.61                                     & 0.10                                                        & 0.77                                     & 0.00                                                        & 0.74                                     & 0.00                                                        & 0.87                                     & 0.00                                                        \\ \cline{2-13} 
                                  & \multirow{3}{*}{Holding}                                                            & A                                                                                   & 0.78                                     & 0.01                                                        & 0.72                                     & 0.01                                                        & 0.69                                     & 0.00                                                        & 0.68                                     & 0.00                                                        & 0.73                                     & 0.00                                                        \\
                                  &                                                                                     & R                                                                                   & 0.64                                     & 0.00                                                        & 0.67                                     & 0.00                                                        & 0.62                                     & 0.03                                                        & 0.73                                     & 0.00                                                        & 0.56                                     & 0.03                                                        \\
                                  &                                                                                     & D                                                                                   & 0.85                                     & 0.01                                                        & 0.78                                     & 0.03                                                        & 0.73                                     & 0.00                                                        & 0.75                                     & 0.00                                                        & 0.66                                     & 0.00                                                        \\ \hline
\multirow{6}{*}{Laying}           & \multirow{3}{*}{Normal}                                                             & A                                                                                   & 0.90                                     & 0.00                                                        & 0.85                                     & 0.03                                                        & 0.69                                     & 0.07                                                        & 0.68                                     & 0.03                                                        & 0.85                                     & 0.03                                                        \\
                                  &                                                                                     & R                                                                                   & 0.89                                     & 0.07                                                        & 0.82                                     & 0.03                                                        & 0.85                                     & 0.00                                                        & 0.84                                     & 0.03                                                        & 0.82                                     & 0.00                                                        \\
                                  &                                                                                     & D                                                                                   & 0.70                                     & 0.00                                                        & 0.60                                     & 0.03                                                        & 0.63                                     & 0.02                                                        & 0.70                                     & 0.00                                                        & 0.74                                     & 0.00                                                        \\ \cline{2-13} 
                                  & \multirow{3}{*}{Holding}                                                            & A                                                                                   & 0.86                                     & 0.00                                                        & 0.82                                     & 0.00                                                        & 0.86                                     & 0.00                                                        & 0.85                                     & 0.00                                                        & 0.82                                     & 0.00                                                        \\
                                  &                                                                                     & R                                                                                   & 0.86                                     & 0.00                                                        & 0.73                                     & 0.00                                                        & 0.74                                     & 0.00                                                        & 0.73                                     & 0.00                                                        & 0.80                                     & 0.00                                                        \\
                                  &                                                                                     & D                                                                                   & 0.88                                     & 0.00                                                        & 0.80                                     & 0.00                                                        & 0.78                                     & 0.00                                                        & 0.60                                     & 0.00                                                        & 0.88                                     & 0.00                                                        \\ \hline
\end{tabular}
}
\end{table*}

\revised{
In this subsection, we compare the experimental results for different postures, breathing patterns, and radar positions. 
And we include all the experimental results of the five points of interest.
Table.~\ref{table: experimental results} presents the experimental results (maximum values of cross-correlations $
{{\rho}}_{rx}$ and differences in maximum frequencies ${{\left( {\Delta f_{rx}} \right)}}_{\text{max}}$) for two postures (sitting and laying) and two breathing patterns (normal breathing and holding breathing), and three radar positions.
}
\revised{By comparing the results in various experimental scenarios, we have the following conclusions:}
\begin{itemize}
    \item The breathing pattern has significant influence on heartbeat detection. \revised{When the subject is under holding breathing pattern, the heartbeat frequencies were accurately estimated, and the error were within $0.03~\text{Hz}$. However, under normal breathing pattern, the errors of estimated heartbeat frequency were from $0~\text{Hz}$ to $0.10~\text{Hz}$ due to the respiratory signal and its harmonics \cite{ren2021vital}. In addition, the radar estimated displacements show more irregularity under normal breathing pattern for the same reasons.}
    \item \revised{The postures and radar positions have no significant influence on the detection results.} This is because in all these scenarios the human subject were static and did not present any body movement. 
\end{itemize}

\subsection{Limitations and Possible Solutions}
As mentioned above, the heartbeat detection results are influenced by the respiratory signal and its harmonics, which has been analyzed in existing studies \cite{li2006experiment,ren2021vital}. 
\revised{
Besides the general cause like respiration signal and its harmonics, another factor, limited elevation resolution, also  influences the detection performance of our MultiVital system.}
 
In our system, the azimuth and elevation resolutions are $1.4^{\circ}$ and $18^\circ$, respectively, which means the cross-range resolution (range resolution in the tangential direction) are $1.22~\text{cm}$ and $15.71~\text{cm}$ for a human subject at the distance of 0.50 m to the radar. 
With these hardware parameters, it is not easy to completely separate the displacements of two points, which are at close azimuth angles, such as point A and point T. 
The displacement of point A estimated by radar, for example, is actually the superposition of the real displacement of point A and that of point T. 
The estimated complex value of the range profile at point A is the linear superposition of the sinc functions of point A and point T, which have different positions, amplitudes, and phases \cite{ren2023tracking}. 
And the phases of the complex values of the range profile are the estimated displacements. 
Although there is an error, the method in this paper is reasonable since the signal at point A dominates the estimation of the displacement of point A \cite{ren2023tracking}. 
There are two possible solutions to this problem. 
The first involves modifications to the system hardware: employing two radars, one positioned horizontally and the other vertically, to achieve high resolution in both azimuth and elevation directions. 
The second solution approaches the problem from an algorithmic perspective, by implementing an angular super-resolution algorithm. 
To completely distinguish the two points A and T at a distance of $0.5~\text{m}$, we need to realize two times of angular super-resolution.


\section{Conclusion}
\label{sec:conclusion}

\begin{figure*}[h]
    \centering
    \includegraphics[width=\textwidth]{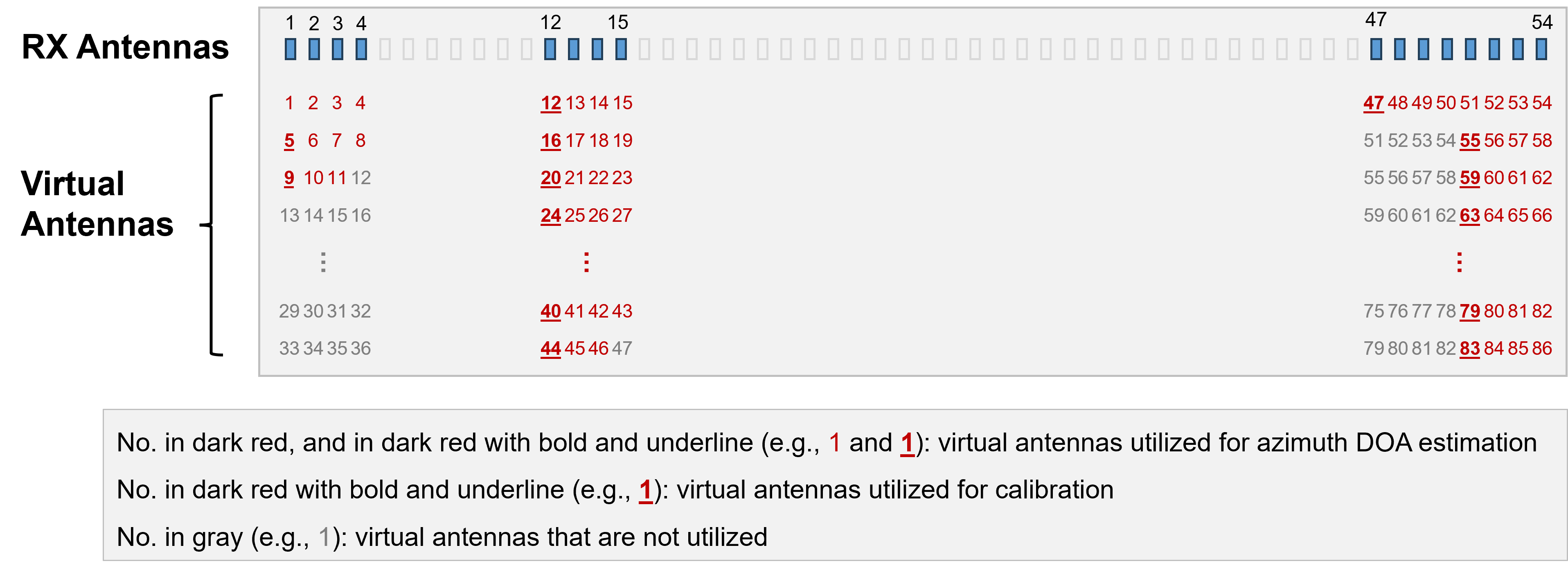}
    \caption{\revised{Virtual antennas selected for azimuth DOA estimation and for calibration. Considering that the azimuth virtual antennas are overlapped and abundant, only 86 of 144 virtual antennas are used to estimate target azimuth. The antennas in dark red are used. There are 20 virtual antennas utilized for phase error compensation.}}
    \label{fig:virtual antennas for calibration}
\end{figure*}

\begin{figure}[h]
    \centering
    \includegraphics[width=0.5\textwidth]{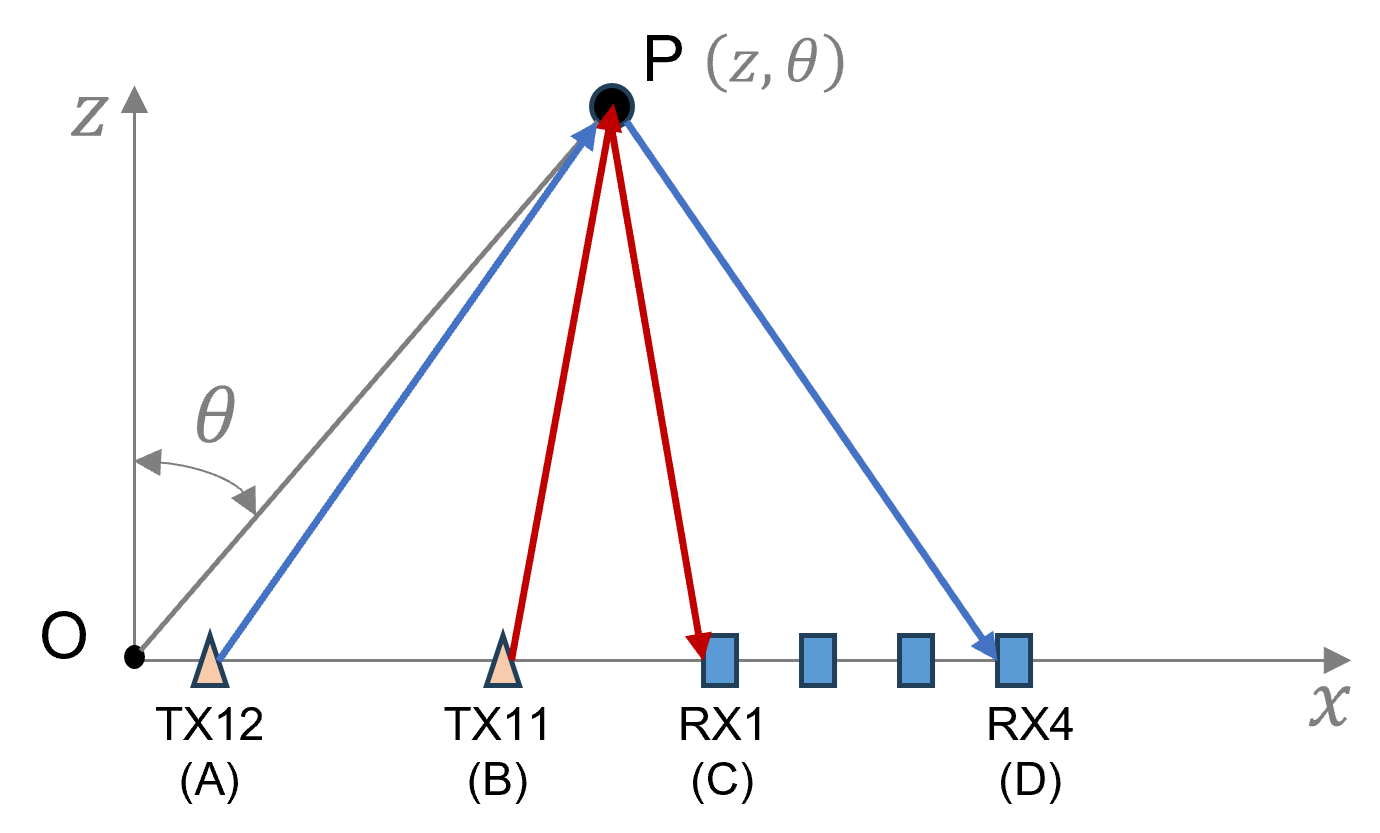}
    \caption{\revised{Near-field geometry for phase error compensation.}}
    \label{fig:phase_error}
\end{figure}

In this study, we present MultiVital, a wireless solution for synchronous monitoring of multi-point vital signs, comprising both hardware and algorithmic parts. The hardware system is built by integrating a mmWave MIMO radar system, five-channel SCG sensors, and one-channel ECG electrodes, and achieve high-precision synchronization across this tri-modal system.
The algorithmic part is a comprehensive signal processing framework, including a radar signal processing module, an SCG calibration module, and a spatial alignment scheme.
As the core of this framework, the radar signal processing module's feasibility and effectiveness are validated through mathematical derivation, simulation, and experiments.
Both simulation and experimental outcomes demonstrate robust performance of our solution in the synchronous monitoring of multi-point vital signs. 
This system can provide dual reference modalities (SCG and ECG) for radar-based multi-point vital sign monitoring, and thus serve for the research of radar-based vital sign monitoring, especially for multi-point applications. 
Using this MultiVital system, the researchers can easily assess the performance of the radar module and refine algorithms accordingly to finally develop a radar-based solution for better human vital sign monitoring. 

This work represents a fundamental contribution to multi-point vital sign monitoring. 
This system enables the precise detection of subtle cardiopulmonary movements in different regions of the human body, providing more accurate and comprehensive information for cardiopulmonary health monitoring.
\revised{This study represents the first effort to differentiate multi-point vital signs on the human chest based on the system’s capacity to achieve this differentiation “physically,” rather than relying solely on computational methods.}
Future research and applications can build on this work, utilizing this tri-modal hardware system and its signal processing framework to explore a variety of application research, such as transformation of the three modality signals and multi-point data fusion.


Although we have made considerable progress by developing such a radar-based multi-point vital sign monitoring system, there are still some limitations of our work. 
\revised{As mentioned above, the accuracy of this solution including the hardware system and signal processing framework suffers from limited elevation resolution.} 
\revised{As future work, we will enhance the radar system through hardware integration of dual mmWave radars for improved azimuth and elevation resolution, and the development of super-resolution algorithms specific to the MultiVital system's application scenarios.}
\revised{Additionally, implementing beamforming at both the transmitting and receiving ends will be considered to further enhance system performance. }

\section*{\revised{Appendix}} \label{sec:appendix}

\begin{figure*}[!t]
 \normalsize
 \revised{
 \begin{align} \label{eq:azimuth FFT}
& \mathcal{F}(\mathbf{x}, l) \\ \nonumber
& = \sum_{i=0}^1 \mathcal{F}_{4i+1:4i+4}(\mathbf{x}, l) e^{-j \frac{2\pi l 4i}{N_{\text{FFT}}}} + \mathcal{F}_{9:11}(\mathbf{x}, l) e^{-j \frac{2\pi l 8}{N_{\text{FFT}}}}  + \sum_{i=3}^{10} \mathcal{F}_{4i:4i+3}(\mathbf{x}, l) e^{-j \frac{2\pi l (4i - 1)}{N_{\text{FFT}}} } + \mathcal{F}_{44:46}(\mathbf{x}, l) e^{-j \frac{2\pi l 43}{N_{\text{FFT}}}} \\ \nonumber
&\quad + \sum_{i=12}^{21} \mathcal{F}_{4i-1:4i+2}(\mathbf{x}, l) e^{-j \frac{2\pi l (4i - 2)}{N_{\text{FFT}}} }
\end{align}
\begin{align} \label{eq:modified azimuth FFT}
& \mathcal{F}_{\text{near}}(\mathbf{x}, l) \\ \nonumber
& = \sum_{i=0}^1 \mathcal{F}_{4i+1:4i+4}(\mathbf{x}, l) e^{-j \frac{2\pi k 4i}{N_{\text{FFT}}}} e^{-j \sum_{h=0}^i \Delta \phi_{4h \rightarrow 4h+1}(z, l)} 
+ \mathcal{F}_{9:11}(\mathbf{x}, l) e^{-j \frac{2\pi k 8}{N_{\text{FFT}}}} e^{-j \sum_{h=0}^2 \Delta \phi_{4h \rightarrow 4h+1}(z, l)} \\ \nonumber
& \quad + \sum_{i=3}^{10} \mathcal{F}_{4i:4i+3}(\mathbf{x}, l) e^{-j \frac{2\pi k (4i - 1)}{N_{\text{FFT}}}} e^{-j \left( \sum_{h=3}^i \Delta \phi_{4h-1 \rightarrow 4h}(z, l) + \sum_{h=0}^2 \Delta \phi_{4h \rightarrow 4h+1}(z, l) \right)} \\ \nonumber
& \quad + \mathcal{F}_{44:46}(\mathbf{x}, l) e^{-j \frac{2\pi k 43}{N_{\text{FFT}}}} e^{-j \left( \sum_{h=3}^{11} \Delta \phi_{4h-1 \rightarrow 4h}(z, l) + \sum_{h=0}^2 \Delta \phi_{4h \rightarrow 4h+1}(z, l) \right)} \\ \nonumber
& \quad + \mathcal{F}_{47:54}(\mathbf{x}, l) e^{-j \frac{2\pi k 46}{N_{\text{FFT}}}} e^{-j \left( \Delta \phi_{46 \rightarrow 47}(z, l) + \sum_{h=3}^{11} \Delta \phi_{4h-1 \rightarrow 4h}(z, l) + \sum_{h=0}^2 \Delta \phi_{4h \rightarrow 4h+1}(z, l) \right)} \\ \nonumber
& \quad + \sum_{i=13}^{21} \mathcal{F}_{4i-1:4i+2}(\mathbf{x}, l) e^{-j \frac{2\pi k (4i - 2)}{N_{\text{FFT}}}} e^{-j \left( \sum_{h=13}^i \Delta \phi_{4h-2 \rightarrow 4h-1}(z, l) + \Delta \phi_{46 \rightarrow 47}(z, l) + \sum_{h=3}^{11} \Delta \phi_{4h-1 \rightarrow 4h}(z, l) + \sum_{h=0}^2 \Delta \phi_{4h \rightarrow 4h+1}(z, l) \right)}.    
\end{align}
}
\hrulefill
\end{figure*}

\subsection*{\revised{Near-field Calibration with Modified Azimuth FFT}}
\subsubsection{\revised{Near-Field Calibration is necessary}}
\revised{
It is assumed in Section~\ref{sec:radar signal processing} that the target is located in the far-field region where the radar waveforms between the target and multiple Tx/Rx antennas are parallel, meaning the phase difference of radar signals for different Tx-Rx pairs depends solely on the target angle. However, in our research scenario, the far-field condition $(R > 2D^2/\lambda)$ is not satisfied.
}

\revised{
To accurately estimate target positions, near-field calibration is required to compensate for the phase error. One insight here is that the most significant phase errors stem from the phase differences between paths from different Tx and Rx antennas.
}

\subsubsection{\revised{Selected Virtual Antennas}}
\revised{
In this study, as shown in Figure~\ref{fig:virtual antennas for calibration}, there are totally 144 virtual antennas formulated by 9 Txs and 16 Rxs, of which 86 ones are used to estimate the target azimuth. In this case, the phase errors correspond to 20 virtual antennas have to be calculated and compensated.}

\subsubsection{\revised{Phase Error Calculation}}
\revised{
To illustrate the computation of phase errors, consider the phase difference between virtual antenna 15 and 16, as the example shown in \ref{fig:phase_error}. If the virtual antennas have uniform spacing of $\lambda/2$, the phase difference under far-field conditions is given by
$ \phi_{\text{far}} = \pi \sin \theta$. 
The phase error is then:
\begin{align}
& \Delta \phi_{15 \rightarrow 16}(z, \theta) \\ \nonumber
& = \phi_{15 \rightarrow 16,\text{near}} - \phi_{\text{far}} \\ \nonumber
& = \frac{2\pi}{\lambda} \left( \left( \overline{\text{BP}} + \overline{\text{CP}} \right) - \left( \overline{\text{AP}} - \overline{\text{DP}} \right) \right) - \pi \sin \theta,
\end{align}
where $\overline{\text{AP}}$, $\overline{\text{BP}}$, $\overline{\text{CP}}$, and $\overline{\text{DP}}$ can be calculated using the Cosine Law as 
\begin{align}
\overline{\text{AP}} = \left( \overline{\text{OP}}^2 + \overline{\text{OA}}^2 - 2 \overline{\text{OP}} \cdot \overline{\text{OA}} \sin \theta \right) \\ \nonumber 
\overline{\text{BP}} = \left( \overline{\text{OP}}^2 + \overline{\text{OB}}^2 - 2 \overline{\text{OP}} \cdot \overline{\text{OB}} \sin \theta \right) \\ \nonumber 
\overline{\text{CP}} = \left( \overline{\text{OP}}^2 + \overline{\text{OC}}^2 - 2 \overline{\text{OP}} \cdot \overline{\text{OC}} \sin \theta \right) \\ \nonumber 
\overline{\text{DP}} = \left( \overline{\text{OP}}^2 + \overline{\text{OD}}^2 - 2 \overline{\text{OP}} \cdot \overline{\text{OD}} \sin \theta \right) 
\end{align}
and $\overline{\text{OP}} = z/\cos \theta$.
}

\subsubsection{\revised{Modified Azimuth FFT}}
\revised{
After calculating all $\Delta \phi(z, \theta)$, the modified azimuth FFT is implemented for azimuth DOA estimation. 
Let $\mathbf{x}$ be a $1 \times 86$ vector corresponding to the selected 86 virtual antennas. The $N_{\text{FFT}}$-point FFT of $\mathbf{x}$ is given by$\mathcal{F}(\mathbf{x}) = \text{FFT}(\mathbf{x}, N_{\text{FFT}})$.
Let \( \mathcal{F}_{i_1:i_2}(\mathbf{x}) \) denote the \( N_{\text{FFT}} \)-point FFT of \( \mathbf{x}(i_1:i_2) \), and let \( \mathcal{F}(\mathbf{x}, l) \) denote the \( l \)-th entry of \( \mathcal{F}(\mathbf{x}) \). In this study, \( \mathcal{F}(\mathbf{x}) \) can be expressed as a combination of 20 terms \( \mathcal{F}_{i_1:i_2}(\mathbf{x}) \) according to the virtual antennas for phase error compensation listed in Fig.~\ref{fig:virtual antennas for calibration}. Specifically, without near-field calibration, this can be represented as (\ref{eq:azimuth FFT}), where \( -\frac{N_{\text{FFT}}}{2} \leq l \leq \frac{N_{\text{FFT}}}{2} - 1 \).
}

\revised{
The modified azimuth FFT is implemented by incorporating near-field calibration which is expressed as (\ref{eq:modified azimuth FFT}), with \( \theta = \arcsin\left( 2l/N_{\text{FFT}} \right) \) and \( \Delta \phi_{0 \rightarrow 1}(z, l) = 0 \).
}


\newpage
\bibliographystyle{IEEEtran} 
\bibliography{mybib}

\newpage
\begin{IEEEbiography}
[{\includegraphics[width=1in,height=1.25in,clip,keepaspectratio]{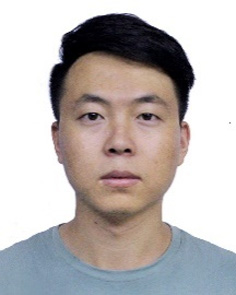}}]
{Wei Ren}
(Member, IEEE) was born in Heilongjiang, China, in 1993. He received the B.S. degree from the Department of Precision Instrument, Tsinghua University, Beijing, China, in 2015, and Ph.D. degree from School of Electronics and Information, Beijing Institute of Technology, China, in 2021. He was also a visiting student with the Department of Electrical Engineering and Computer Science, The University of Tennessee, Knoxville, TN, USA from 2019 to 2020. 

He is currently working as a Postdoc Fellow with the Department of Computing, The Hong Kong Polytechnic University, Hong Kong. His research interests include wireless human sensing, radar waveform design, radar signal processing, and machine learning.
\end{IEEEbiography}

\begin{IEEEbiography}
[{\includegraphics[width=1in,height=1.25in,clip,keepaspectratio]{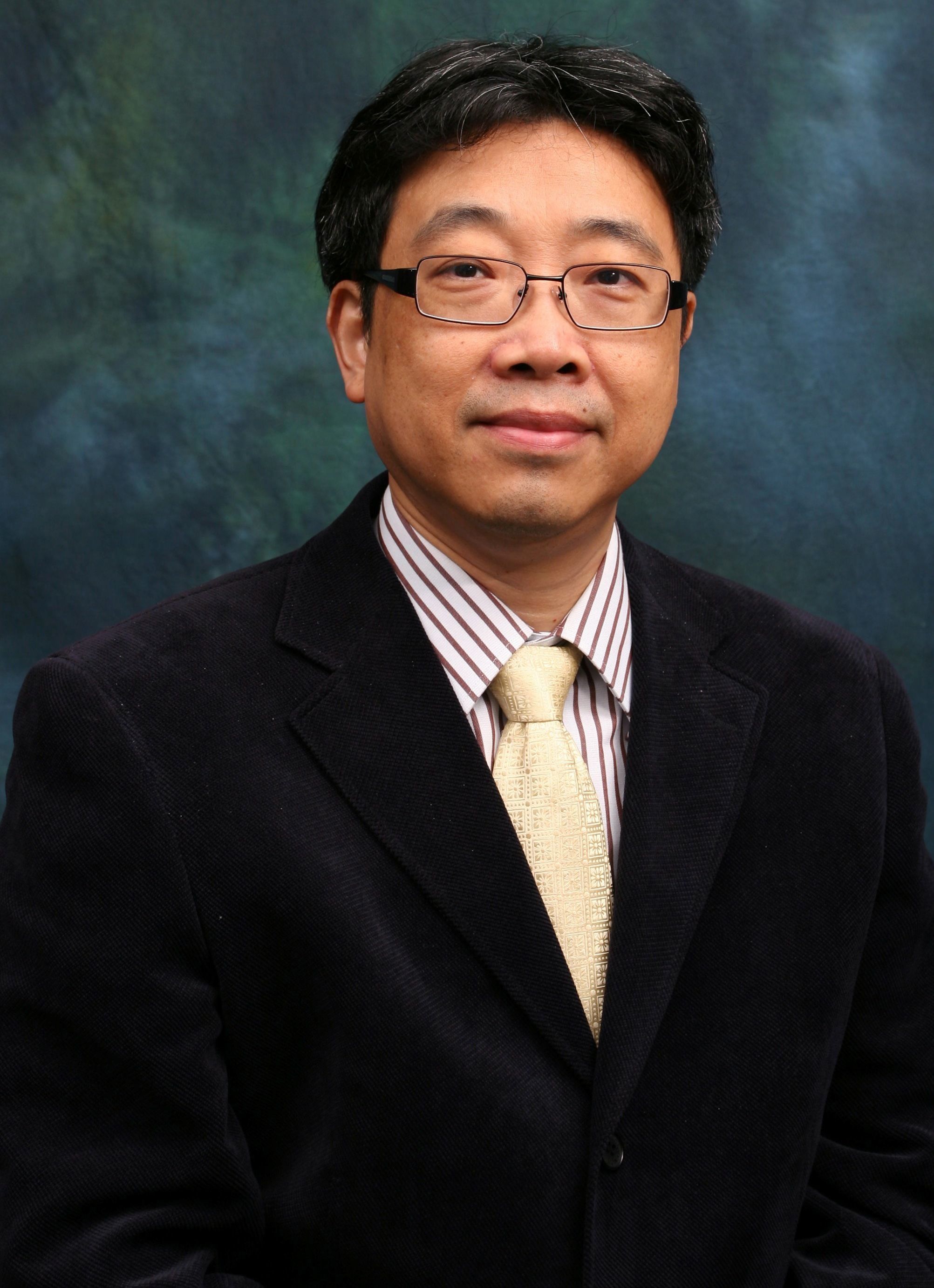}}]
{Jiannong Cao}
(Fellow, IEEE) received his Ph.D. degree in computer science from Washington State University, Pullman, WA, USA, in 1990. 

He is currently the Otto Poon Charitable Foundation Professor in Data Science and the Chair Professor of Distributed and Mobile Computing in the Department of Computing at PolyU. He is also the Dean of Graduate School, the director of the Research Institute for Artificial Intelligence of Things, and the director of the Internet and Mobile Computing Lab.
Prof. Cao’s research interests include Distributed Systems and Blockchain, Big data and Machine learning, Wireless Sensing and Networking, and Mobile Cloud and Edge Computing.
\end{IEEEbiography}

\begin{IEEEbiography}
[{\includegraphics[width=1in,height=1.25in,clip,keepaspectratio]{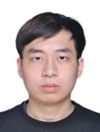}}]
{Huansheng Yi} received the B.E. degrees in internet of things engineering from LuoYang Institute of Science and Technology in 2023, Henan, China.

He is currently serving as a RA in the BME department, FMMU, Xi’an. Since 2019, He majored in internet of things and embedded technology, school of computer and information engineering, Luoyang, Henan, China. He won the third prize in the ICAN Innovation Contest(final), 2021. He won the second prize of China Undergraduate Computer Design Competition, 2022. He won the third prize of China Undergraduate Service Outsourcing Innovation and Entrepreneurship Competition, 2022. He also awarded scholarship for many times in college. 
\end{IEEEbiography}

\begin{IEEEbiography}
[{\includegraphics[width=1in,height=1.25in,clip,keepaspectratio]{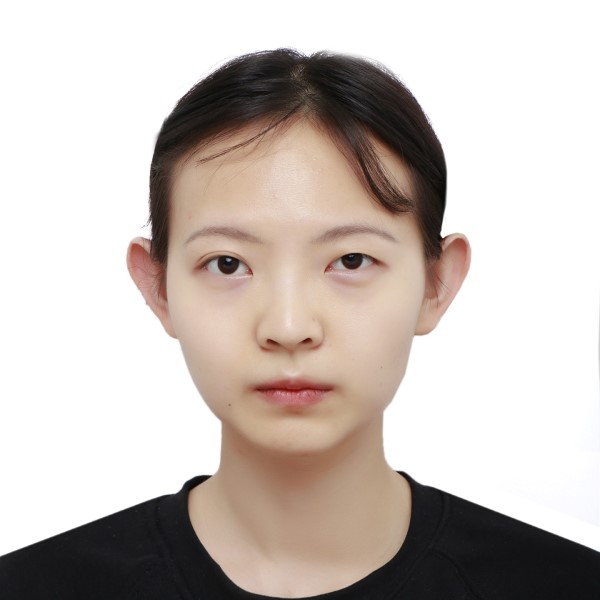}}]
{Kaiyue Hou} received the B.S. degree in communication engineering from Jilin University, China, in 2019. She received the M.S. degree in the Department of information and electronics, Beijing Institute of Technology, China, in 2022.

She is currently pursuing the Ph.D. degree with the Department of Electrical and Electronic Engineering, The Hong Kong Polytechnic University. Her research interests include optimization and integrated sensing and communication.
\end{IEEEbiography}

\begin{IEEEbiography}
[{\includegraphics[width=1in,height=1.25in,clip,keepaspectratio]{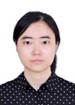}}]
{Miaoyang Hu} received a MBBS and Ph.D. degrees in clinical medicine (eight-year program) in 2014 and 2016 from Wuhan University , Wuhan, China.

Since 2016, she has been an attending physician at department of cardiology, Xijing Hospital, Fourth Military 
Medical University (FMMU) and engaged in clinical cardiac electrophysiology. She has rich clinical practical experience in the diagnosis and interventional treatment of various types of tachyarrhythmias and bradycardiac arrhythmias. In July 2023, she went to study CRT implant and programming optimization in Westmead Hospital, Sydney, NSW. In October 2023, she  went to study cardiac electrophysiology in Nanjing Medical University, Nanjing, China. She has published over 10 peer-reviewed journal and conference papers. Her research interest is cardiac electrophysiology.
\end{IEEEbiography}

\begin{IEEEbiography}
[{\includegraphics[width=1in,height=1.25in,clip,keepaspectratio]{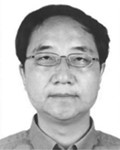}}]
{Jianqi Wang} was born in Xi’an, Shaanxi, China, in 1962. He received the B.E. degree in information and control engineering from Xi’an Jiaotong University, Xi’an, in 1984; the M.E. degree in electronic engineering from the National University of Defense Technology, Changsha, China, in 1990; and the Ph.D. degree from the Key Laboratory of the Ministry of Education of China, Xi’an Jiaotong University, in 2006.

Since 1990, he has taught in the School of Biomedical Engineering, Fourth Military Medical University, Xi’an. He is currently serving as a Professor with and the Director of the Department of Electronics. He pioneered radar-based human being detection in China in 1998 and has published more than 100 articles on the technology. His research interest is bio-radar technology, including signal processing, human being detection, and imaging.
\end{IEEEbiography}

\begin{IEEEbiography}
[{\includegraphics[width=1in,height=1.25in,clip,keepaspectratio]{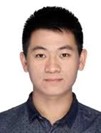}}]
{Fugui Qi} received a B.E. and Ph.D. degrees in bio-medical engineering in 2014 and 2020 from the Institute of Bio-medical engineering (BME), Fourth Military Medical University (FMMU), Xi’an, China.

Since 2020, he has been a Lecturer with the Bio-radar and Signal Processing Laboratory, Institute of Bio-medical engineering, FMMU. During 2019-2020, He is a visiting scholar at the University of Tennessee, Knoxville, USA.  He won the Excellent  Doctorial Dissertations of  Shaanxi Province in 2022, and was selected for the Young Elite Scientists Sponsorship Program of Shaanxi Provincial Association for Science and Technology in 2023. He has published over 40 peer-reviewed journal and conference papers. His research interests include Bio-radar-based medical signal processing, through-wall human motion recognition, and non-contact intelligent life sensing.
\end{IEEEbiography}

\end{document}